\documentclass[5p]{elsarticle}

\usepackage{subfigure}
\usepackage{color}

\usepackage{lineno,hyperref}

\journal{Nuclear Instruments and Methods in Physics Research Section A}

\begin{document}

\begin{frontmatter}

\title{Subnanosecond Time-to-Digital Converter Implemented in a Kintex-7 FPGA}

\author[address_nagoya]{Y.~Sano}
\author[address_nagoya]{Y.~Horii\corref{correspondingauthor}}
\cortext[correspondingauthor]{Corresponding author}
\ead{yhorii@hepl.phys.nagoya-u.ac.jp}
\author[address_kek]{M.~Ikeno}
\author[address_kek]{O.~Sasaki}
\author[address_nagoya]{M.~Tomoto}
\author[address_kek]{T.~Uchida}

\address[address_nagoya]{Nagoya University, Chikusa-ku, Nagoya 464-8602, Japan}
\address[address_kek]{High Energy Accelerator Research Organisation (KEK), Oho, Tsukuba 305-0801, Japan}

\begin{abstract}
Time-to-digital converters (TDCs) are used in various fields,
including high-energy physics.
One advantage of implementing TDCs in field-programmable gate arrays (FPGAs)
is the flexibility on the modification of the logics,
which is useful to cope with the changes in the experimental conditions.
Recent FPGAs make it possible
to implement TDCs with a time resolution less than 10~ps.
On the other hand, various drift chambers require a time resolution of $\cal O$(0.1)~ns,
and a simple and easy-to-implement TDC is useful for a robust operation.
Herein an eight-channel TDC with a variable bin size down to 0.28~ns
is implemented in a Xilinx Kintex-7 FPGA and tested.
The TDC is based on a multisampling scheme with quad phase clocks
synchronised with an external reference clock.
Calibration of the bin size is unnecessary
if a stable reference clock is available,
which is common in high-energy physics experiments.
Depending on the channel,
the standard deviation of the differential nonlinearity for a 0.28 ns bin size is 0.13--0.31.
The performance has a negligible dependence on the temperature.
The power consumption and the potential to extend the number of channels are also discussed.
\end{abstract}


\end{frontmatter}


\section{Introduction}

Time-to-digital converters (TDCs) are used
in various fields, including high-energy physics.
For example, in the ATLAS experiment~\cite{ATLAS}, the muon momentum is measured 
with monitored drift tube (MDT) chambers based on TDCs with a bin size of 0.78~ns~\cite{AMT}.
They are implemented in application-specific integrated circuits. 

Recent moderate field-programmable gate arrays (FPGAs) make it possible
to implement TDCs with a nanosecond or smaller bin size~\cite{CDC, TDC_10ps}.
Implementation of TDCs in FPGAs provides the flexibility
to deal with the changes in the experimental conditions.
The present TDC for the MDT chambers includes
a transmitter with fixed bandwidth of 80 Mbit/s and
has limitations on the readout for increased luminosity~\cite{ATLAS_SD}.
On the other hand, FPGAs can in general manage the bandwidth change.
In addition to the flexibility,
FPGAs have advantages on the availability of highly reliable circuits,
e.g. the clock managers and the data transceivers~\cite{Xilinx_FPGA, Microsemi_FPGA}.


In this study, an eight-channel TDC with a variable bin size down to 0.28~ns
is implemented in a Xilinx Kintex-7 FPGA~\cite{Kintex7} and tested.
The type of the employed FPGA is XC7K325T-2FFG900, where the speed grade is -2.
The TDC is based on multi-phase clocks
managed by a highly reliable clock manager integrated in the FPGA.
Calibration of the bin size is unnecessary
if a stable reference clock is available,
which is common in high-energy physics experiments.
The logic is simple and easy-to-implement. 

This study extends the demonstration described in~\cite{JINST}
by further analysing and interpreting the data
as well as measuring the power consumption.
Additionally, the potential to extend the number of channels to 256 is discussed.
This study does not include the measurements of the radiation tolerance of the FPGA,
for which some results are available in \cite{RadTolerance}.

\section{Design of the TDC}

\begin{figure}[tbp] 
\centering
\includegraphics[width=0.48\textwidth]{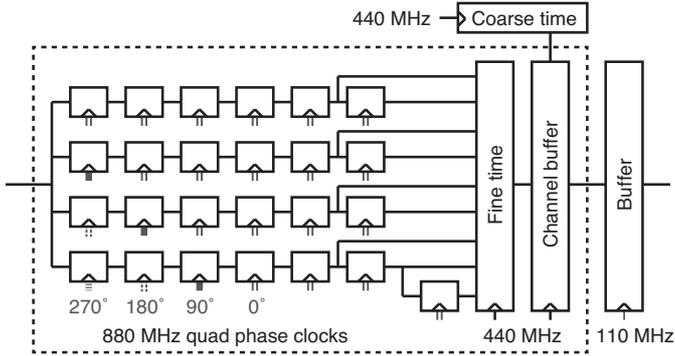}
\caption{Block diagram of the TDC.
The input signal is divided into four
and detected by the D-type flip-flops based on four different clock phases.
The locations of the four D-type flip-flops are
constrained to control the difference between the divided signal paths.
The vertical double lines, thick lines, double dashed lines, and dashed thick line
correspond to the clock phases 0$^\circ$, 90$^\circ$, 180$^\circ$, and 270$^\circ$, respectively.
The outputs from the four D-type flip-flops are aligned step by step
by the chains of additional D-type flip-flops.
The fine time counter extracts a three-bit time count from the pattern of the outputs from the chains.
The fine time count is combined with the data from a fourteen-bit coarse time counter,
and stored in a channel buffer.
Each channel has a circuit inside the square of the dashed line.
The channel buffers for all channels are scanned,
and the data is transferred to a buffer. A three-bit channel identifier is attached.
The data in the buffer are read out one by one.
}
\label{fig:TDC}
\end{figure}

\begin{figure}[tbp] 
\centering
\includegraphics[width=0.48\textwidth]{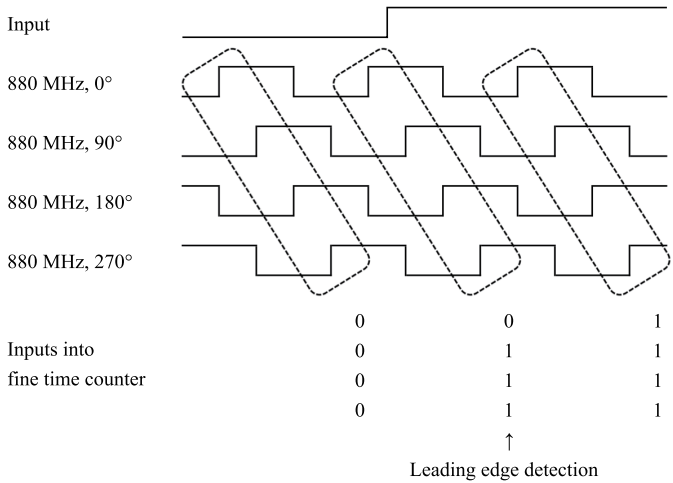}
\caption{Timing chart of an example input and the quad phase clocks,
as well as the resulting inputs into the fine time counter.
The outputs from the four D-type flip-flops corresponding to the four clock phases
are transferred to the fine time counter
after a step-by-step timing alignment by the chains of the D-type flip-flops.
The fine time counter extracts a three-bit time count from the pattern of the four numbers.
In this example, a leading edge is detected and the timing is extracted from the pattern ``0111''.
}
\label{fig:signal_chart}
\end{figure}

Figure~\ref{fig:TDC} shows a block diagram of the TDC.
It is based on a multisampling scheme with quad phase clocks~\cite{TDC1, TDC2, TDC3}.
The input signal is divided into four, and detected by four D-type flip-flops.
The outputs from the D-type flip-flops are aligned step by step
in the chains of additional D-type flip-flops,
where the metastable states are suppressed.
From the pattern of the outputs from the chains, a three-bit fine time count is provided.
The timing of both the leading and trailing edges is extracted.
Figure~\ref{fig:signal_chart} shows a timing chart of an example input signal and the quad phase clocks.

The three-bit time count and a one-bit identifier of the leading and trailing edges
are combined with the data from a fourteen-bit coarse time counter,
and stored in a channel buffer.
The channel buffers for all channels are scanned,
and the data is transferred to a buffer. A three-bit channel identifier is attached.
There is no selection on the data by the trigger, and all the data are simply read out one by one.

The difference in the lengths of the divided input signal paths (see Figure~\ref{fig:TDC})
is crucial for the TDC performance.
The locations of the first four D-type flip-flops in each channel
are constrained so that they are close to each other.

The quad phase clocks are produced by the clock manager of the Kintex-7 FPGA.
They are synchronised with the external reference clock.
The maximum clock frequency supported for the FPGA employed in this study is 933~MHz.
The quad phase clock frequency and the TDC bin size can be varied
by changing the external reference clock frequency.
An external reference clock with a 110~MHz frequency
corresponds to that of the quad phase clocks of 880~MHz and a bin size of 0.28~ns.

The dynamic range for a bin size of 0.28~ns is 37~$\mu$s.
The dynamic range can be extended by changing the number of bits for the coarse time counter.
A minimal latency from the signal input to the buffer output is 0.21~$\mu$s,
while the latency depends on the rate of the signal input.

\section{Demonstrator and Test Setup}

\begin{figure}[tbp] 
\centering
\includegraphics[width=0.48\textwidth]{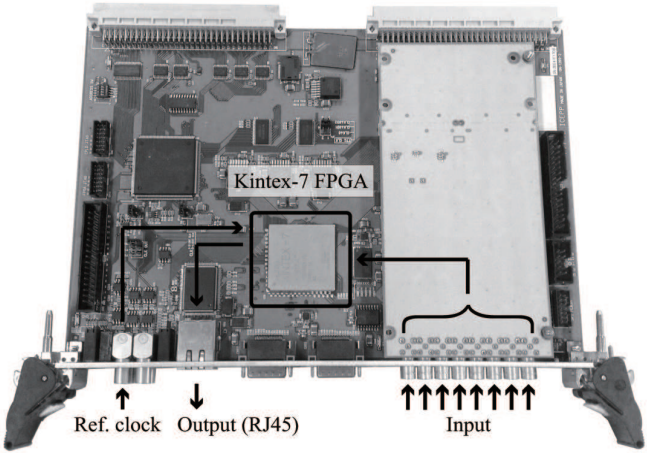}
\caption{Picture of the demonstrator.
The NIM input from the LEMO connectors on the daughter card
is converted into 3.3~V LVCMOS signal and transferred to the FPGA on the motherboard.
The system clock is provided from a LEMO connector on the motherboard.
The output from the FPGA is read through a gigabit Ethernet.}
\label{fig:Demonstrator}
\end{figure}

Figure~\ref{fig:Demonstrator} shows a picture of the demonstrator,
which consists of a motherboard and a daughter card.
The motherboard is compatible with the versatile backplane bus standard VME~\cite{VME}.
It has a Kintex-7 FPGA (type is XC7K325T-2FFG900).
The system clock is provided from a LEMO coaxial connector~\cite{LEMO}.
The output from the FPGA is read through a gigabit Ethernet
with a Transmission Control Protocol processor implemented in the FPGA (SiTCP)~\cite{SiTCP}.
The daughter card has eight LEMO coaxial connectors as signal inputs.
The input compatible with the NIM fast negative signal~\cite{NIM}
is converted into a 3.3~V LVCMOS signal~\cite{LVCMOS}
with Texas Instruments SN65LVDS348PW~\cite{Texas},
and transferred to the FPGA on the motherboard.

The signal and external reference clocks are
from the pulse generator Agilent 81150A~\cite{Agilent}.
The standard deviation of the time difference between the leading edges
of the two synchronised pulses from the pulse generator is 30~ps.
The performances of bin sizes of 0.78~ns, 0.39~ns, and 0.28~ns,
which correspond to the external reference clock frequencies
of 40~MHz, 80~MHz, and 110~MHz, respectively, were evaluated.
Unless otherwise specified, the results for the 0.28-ns bin size are shown.
Although the leading edges were evaluated,
the trailing edges should produce similar results
since common clocks and D-type flip-flops are used.

\section{Differential Nonlinearity}

\begin{figure*}[tbp]
  \centering
  \subfigure[Channel 1.]{
    \includegraphics[width=0.40\textwidth]{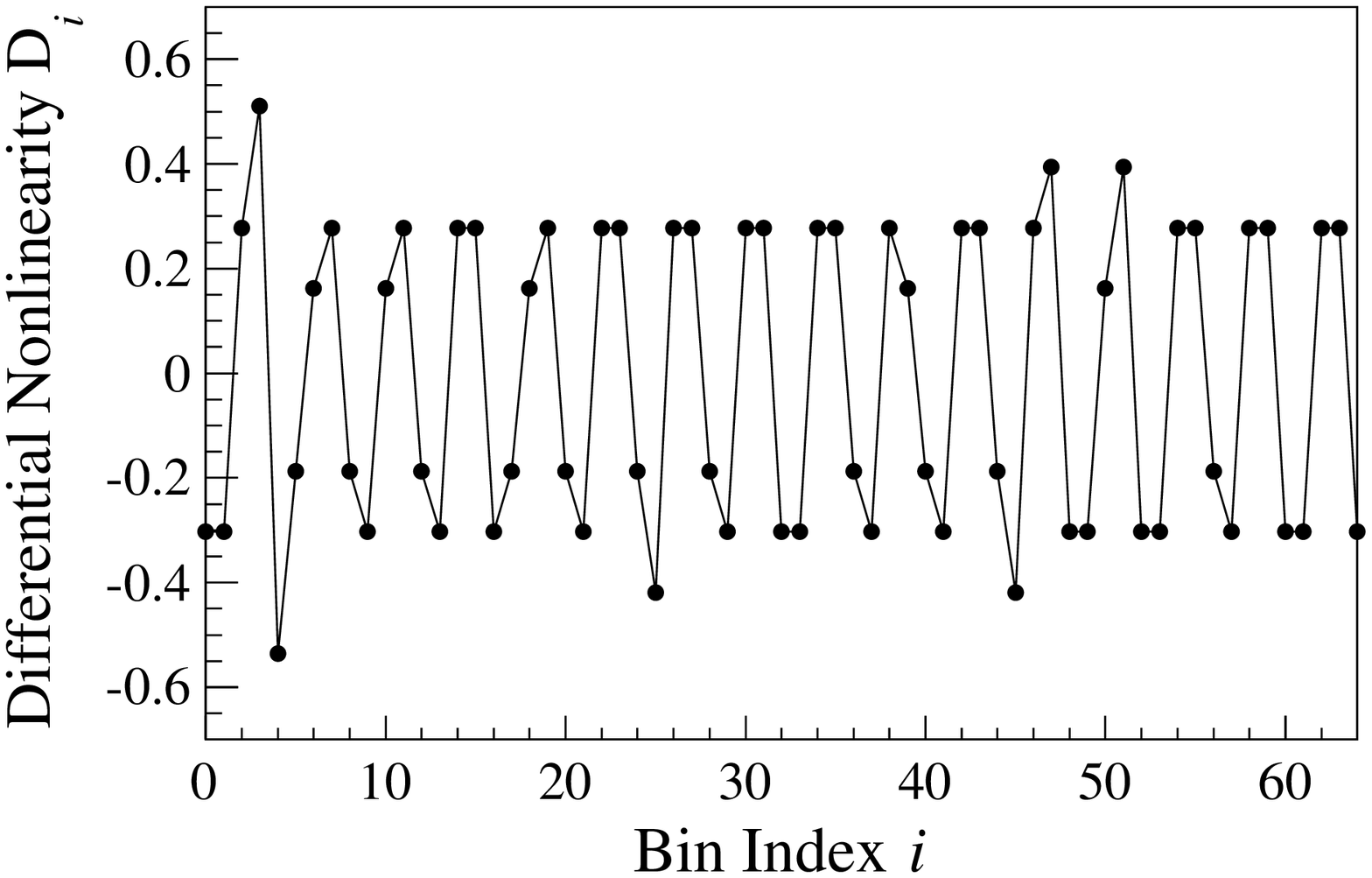}
  \label{fig:label_(a)}}
  \hspace{5mm}
  \subfigure[Channel 2.]{
    \includegraphics[width=0.40\textwidth]{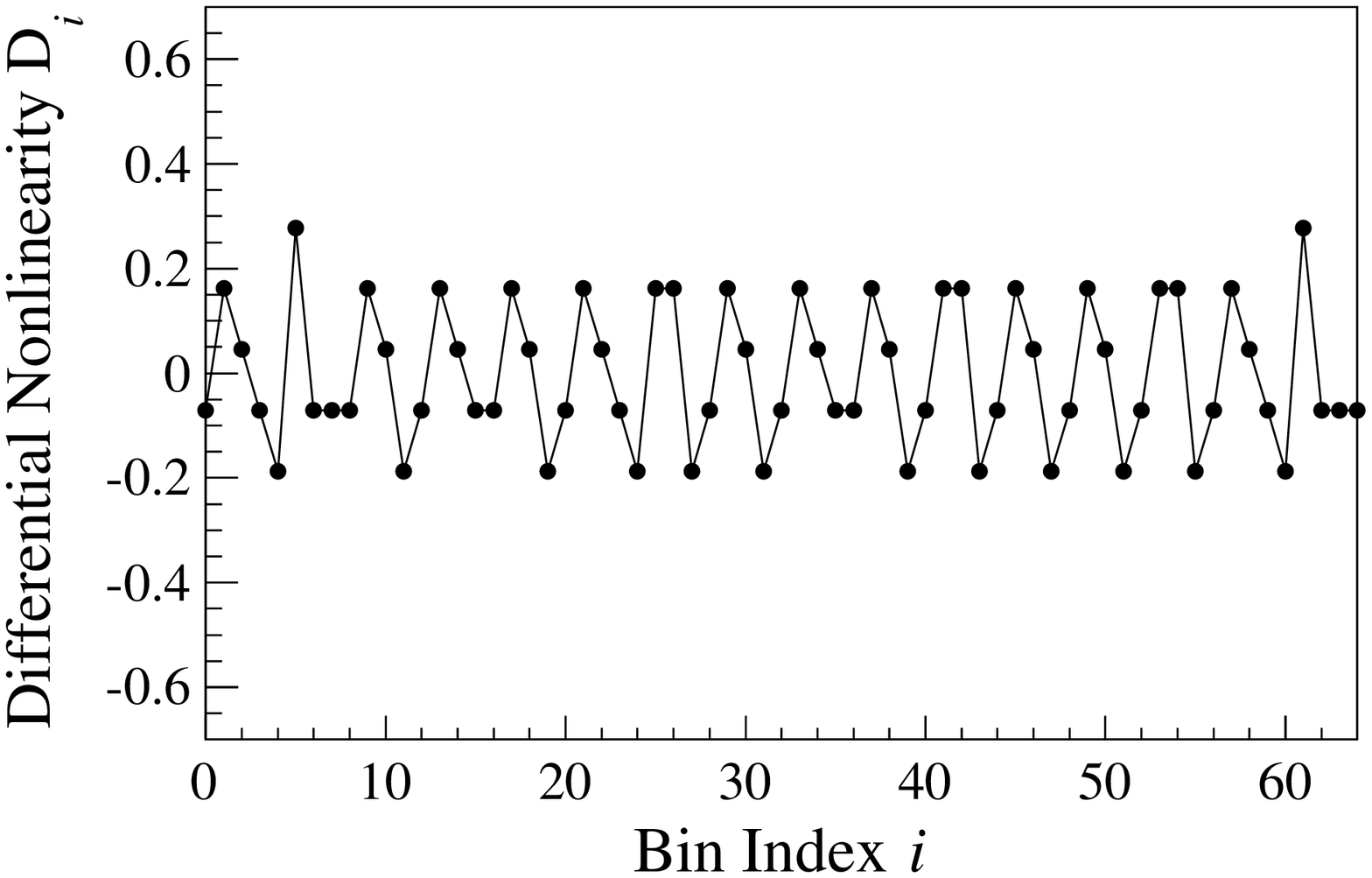}
  \label{fig:label_(b)}}\\
  \vspace{-2mm}
    \subfigure[Channel 3.]{
    \includegraphics[width=0.40\textwidth]{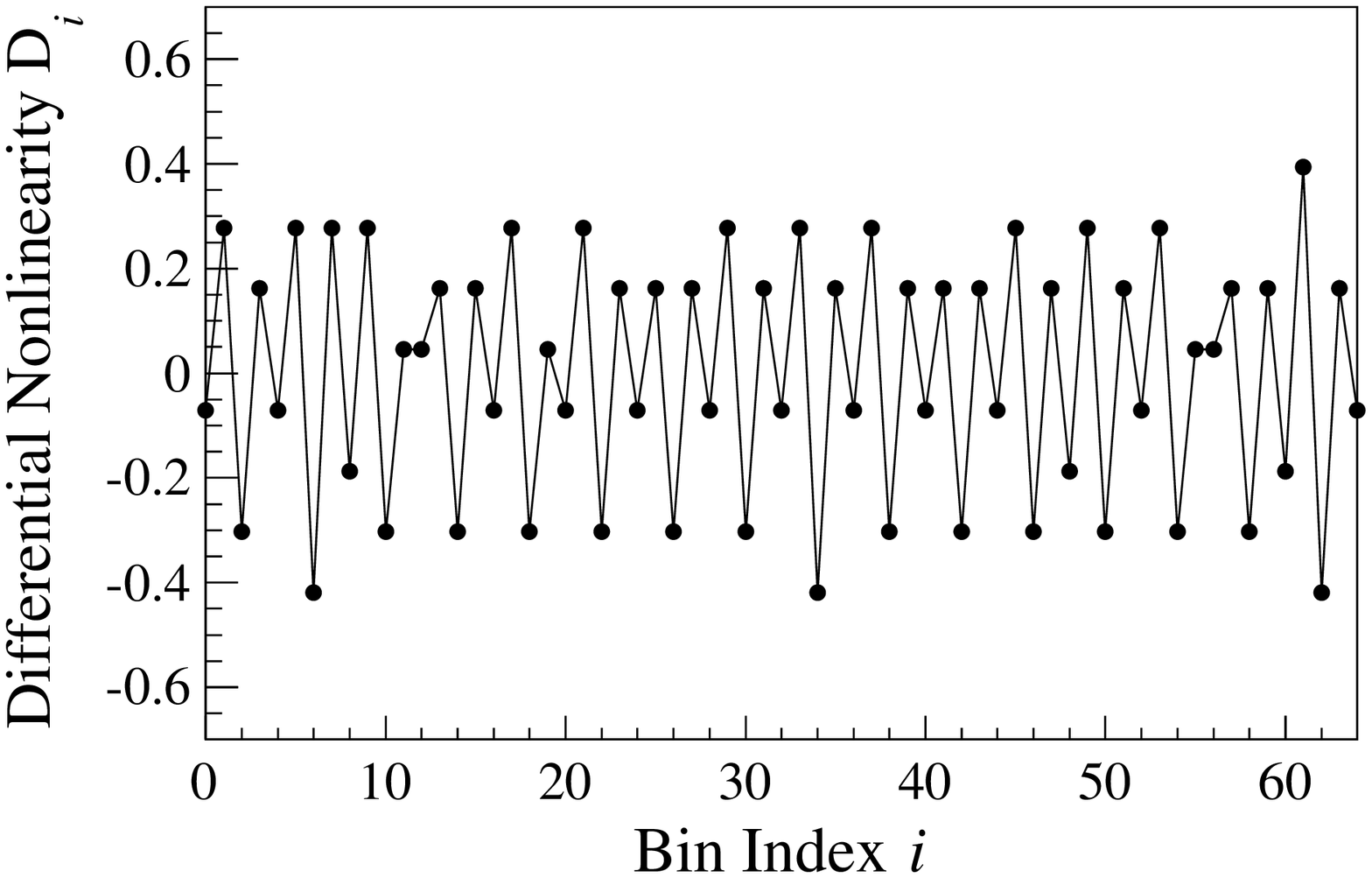}
  \label{fig:label_(c)}}
  \hspace{5mm}
    \subfigure[Channel 4.]{
    \includegraphics[width=0.40\textwidth]{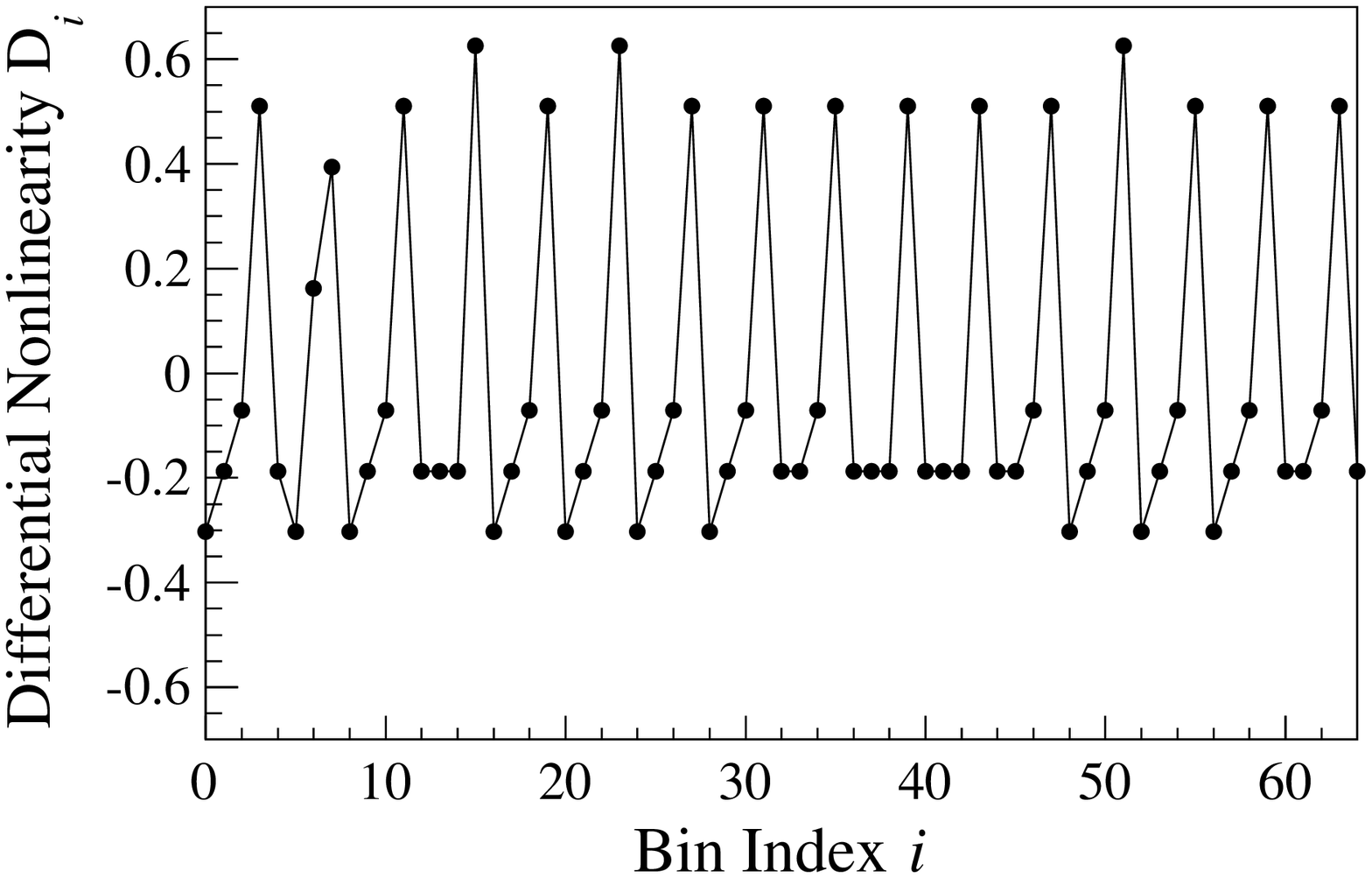}
  \label{fig:label_(d)}}\\
  \vspace{-2mm}
    \subfigure[Channel 5.]{
    \includegraphics[width=0.40\textwidth]{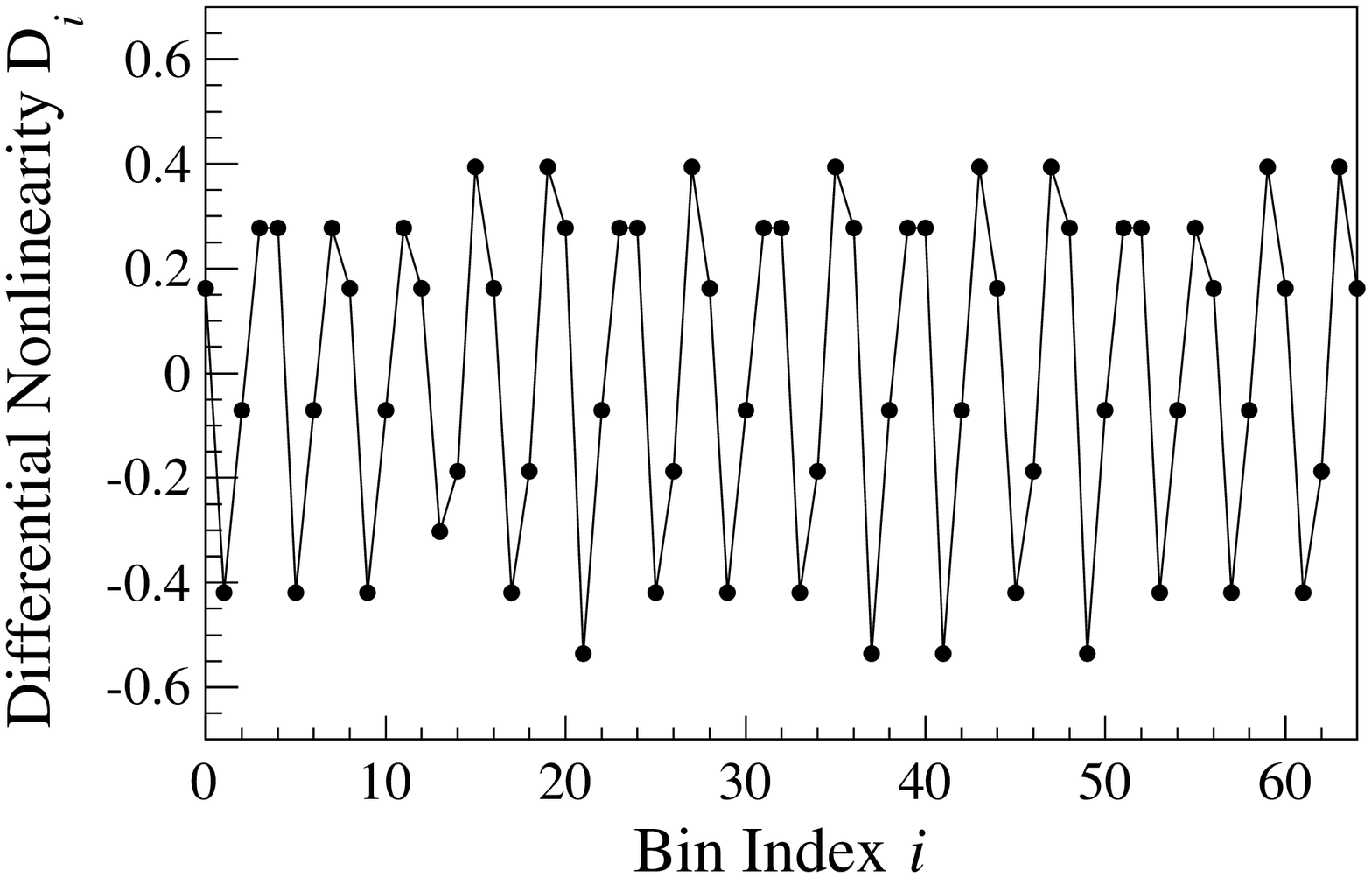}
  \label{fig:label_(e)}}
  \hspace{5mm}
    \subfigure[Channel 6.]{
    \includegraphics[width=0.40\textwidth]{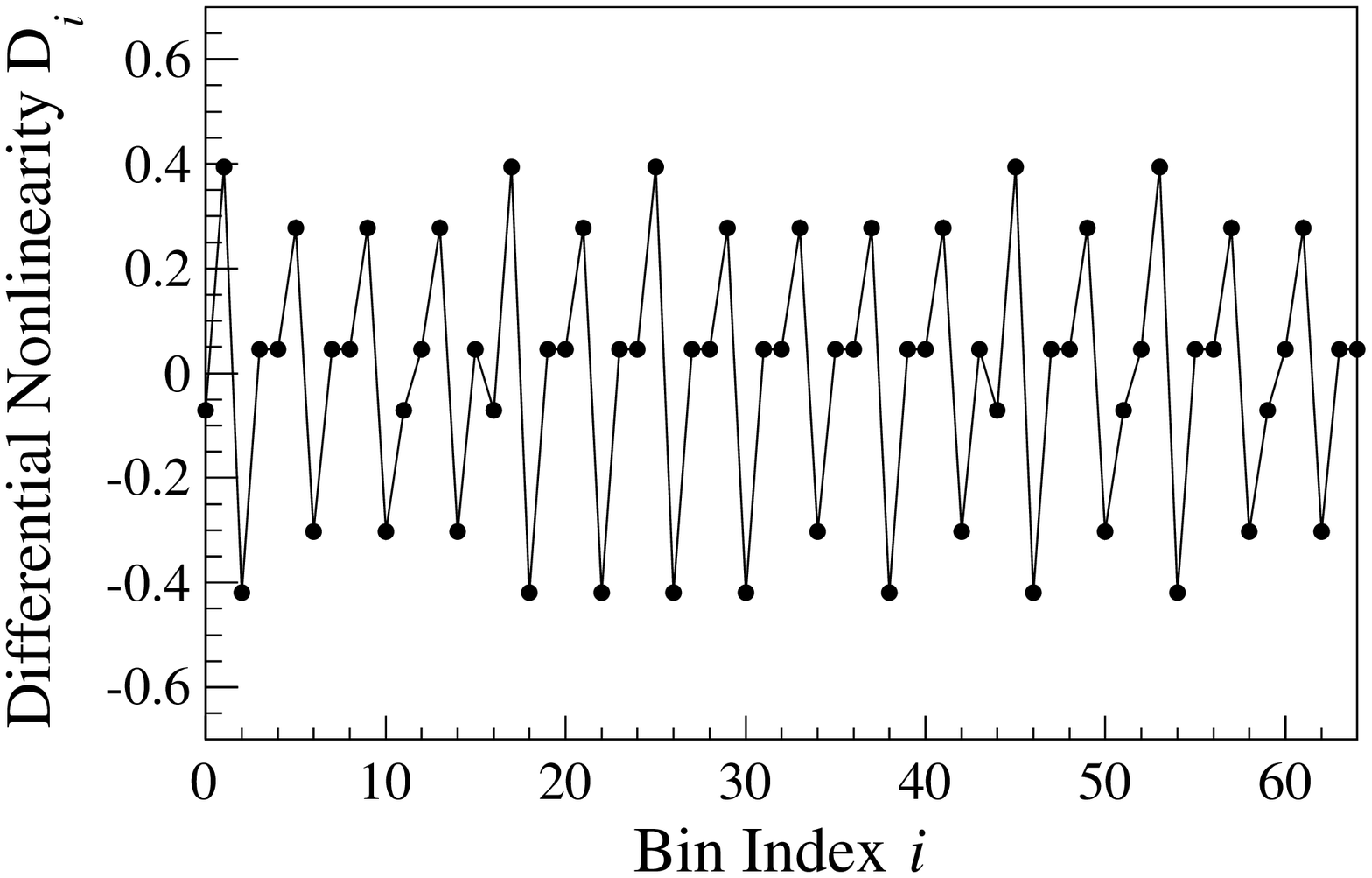}
  \label{fig:label_(f)}}\\
  \vspace{-2mm}
    \subfigure[Channel 7.]{
    \includegraphics[width=0.40\textwidth]{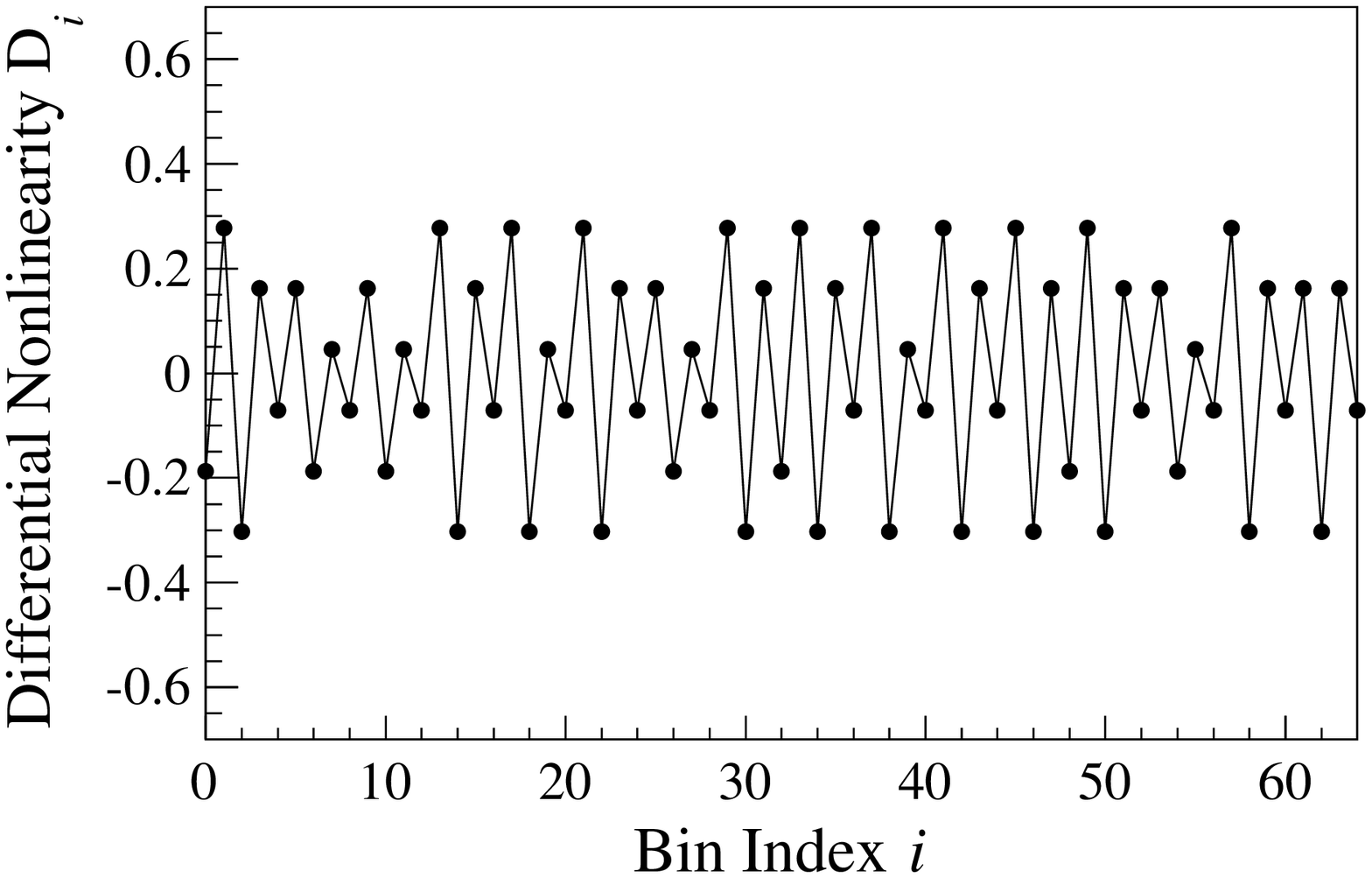}
  \label{fig:label_(g)}}
  \hspace{5mm}
    \subfigure[Channel 8.]{
    \includegraphics[width=0.40\textwidth]{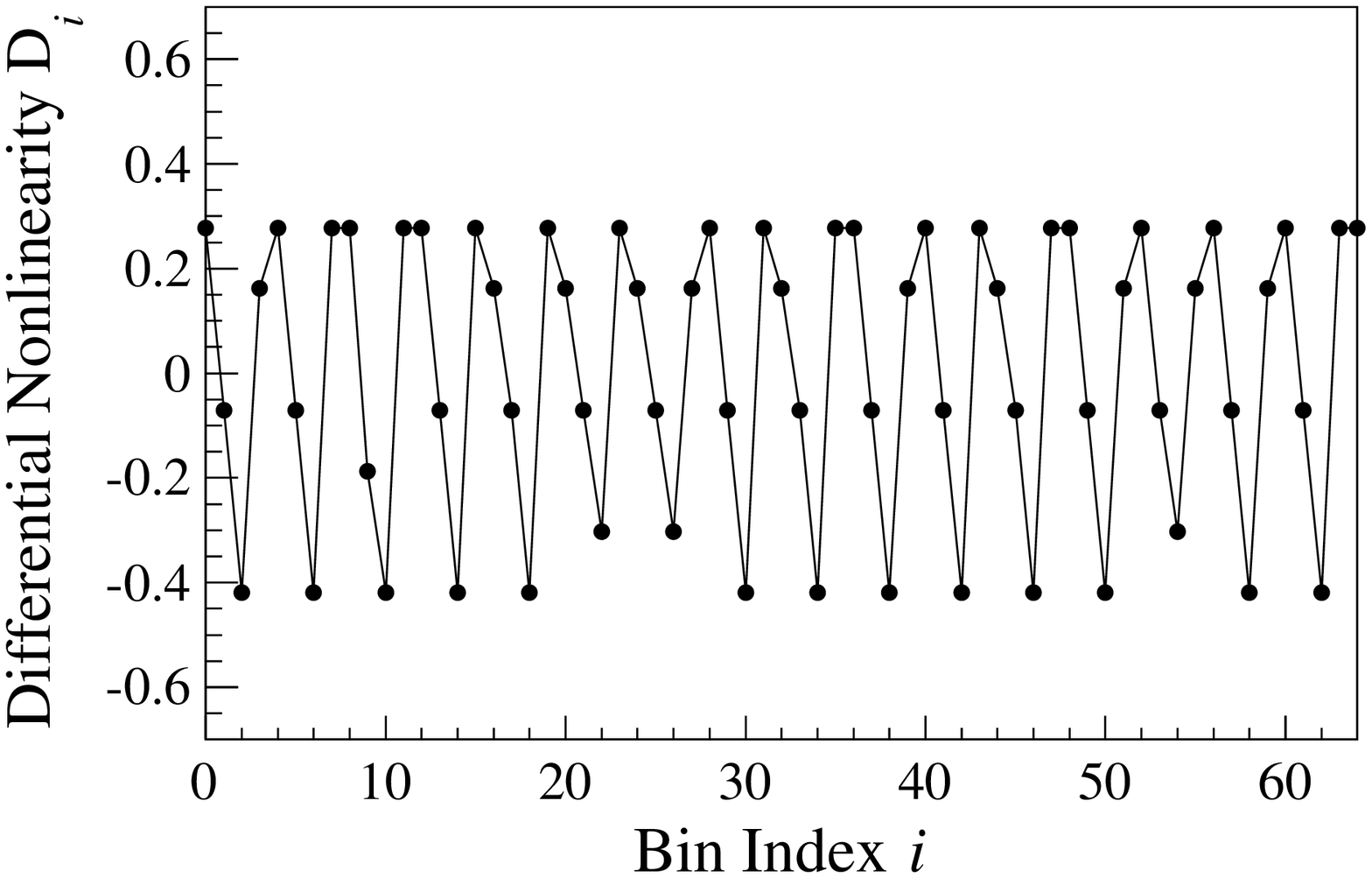}
  \label{fig:label_(h)}}
  \caption{Result of the measurement of the differential nonlinearity $D_i$.
  The values are shown depending on the bin index $i$ for all the implemented channels 1--8.}
  \label{fig:DNL}
\end{figure*}

The differential nonlinearity $D_i$ is defined by
\begin{equation}
D_i = \frac{t_i-t_{\rm bin}}{t_{\rm bin}},
\end{equation}
where $i$ ($i = 0,~1,~2,~...$) is the bin index,
$t_i$ is the bin size for bin $i$,
and $t_{\rm bin}$ is the designed bin size.
To evaluate $D_i$, the time difference of the leading edges
between the signal and external reference clocks was measured.
Multiple bins were investigated
by scanning the time difference between the signal and external reference clocks
with a 33-ps step size (Figure~\ref{fig:DNL}).
A periodic structure with a cycle of four bins was found.
The maximum magnitude of the measured $D_i$ is 0.6.

\begin{figure}[tbp] 
\centering
\includegraphics[width=0.48\textwidth]{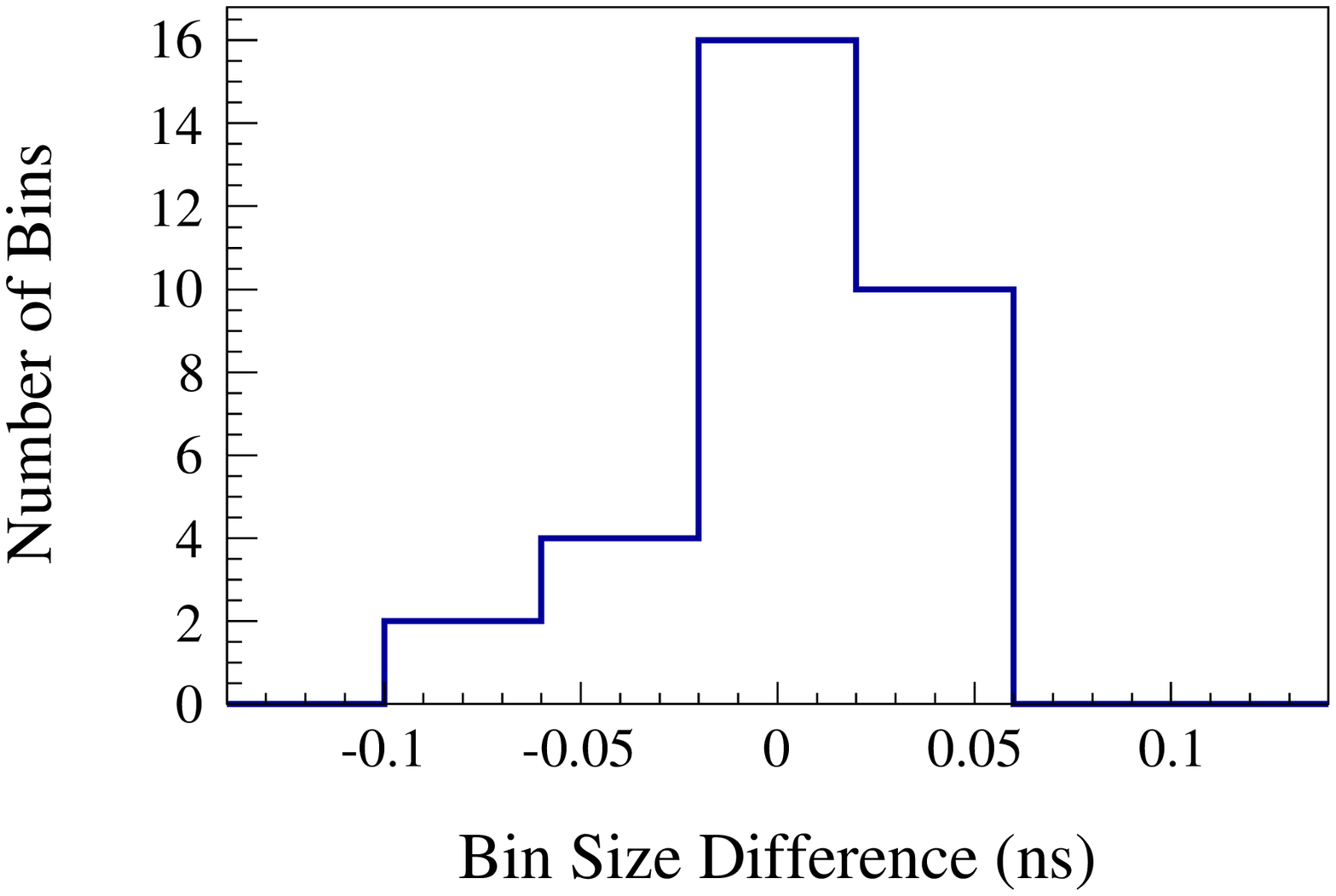}
\caption{Histogram of the difference between the measured and calculated bin sizes
for all 32 quad clock phases of the 8 channels. The magnitude of the difference is
less than 0.1~ns for all entries.}
\label{fig:BIN_size}
\end{figure}

The delay $\Delta_j$ before the first D-type flip-flop
for each divided input signal path $j$ ($j = 0,~1,~2,~3$)
was estimated by a Xilinx Vivado design tool.
Based on the estimated delay,
the bin size $t_j$ for each quad clock phase $j$ was calculated as
\begin{equation}
t_j = t_{\rm bin} + \Delta_j - \Delta_{j+1}.
\label{eq:bin_size}
\end{equation}
A cyclic relation of $\Delta_4 = \Delta_0$ was used
to calculate $t_3$.
Table~\ref{tab:DNL_bin_110MHz} shows the measured and calculated bin sizes for a channel.
Figure~\ref{fig:BIN_size} shows the distribution of the difference
between the measured and calculated bin sizes for all 32 quad clock phases of the 8 channels.
The difference is O(0.01)~ns,
implying that the main source of the periodic structure in Figure~\ref{fig:DNL}
is the difference in the divided input signal paths.

\begin{table}[tbp]
\small
\caption{Relation between the calculated and measured bin sizes for a channel.
The calculated bin sizes are obtained from the input signal delay with Eq.~(\ref{eq:bin_size}).}
\smallskip
\centering
\begin{tabular}{|c|c|c|c|}
\hline
 Phase $j$ & Input signal & Calculated & Measured\\
 & delay $\Delta_j$ [ns] & bin size [ns] & bin size [ns]\\ \hline \hline
 0 & 4.62 & 0.20 & 0.21 \\ \hline
 1 & 4.70 & 0.20 & 0.20 \\ \hline
 2 & 4.79 & 0.34 & 0.36 \\ \hline
 3 & 4.73 & 0.39 & 0.37 \\ \hline
\end{tabular}
\label{tab:DNL_bin_110MHz}
\end{table}

\begin{figure}[tbp] 
\centering
\includegraphics[width=0.48\textwidth]{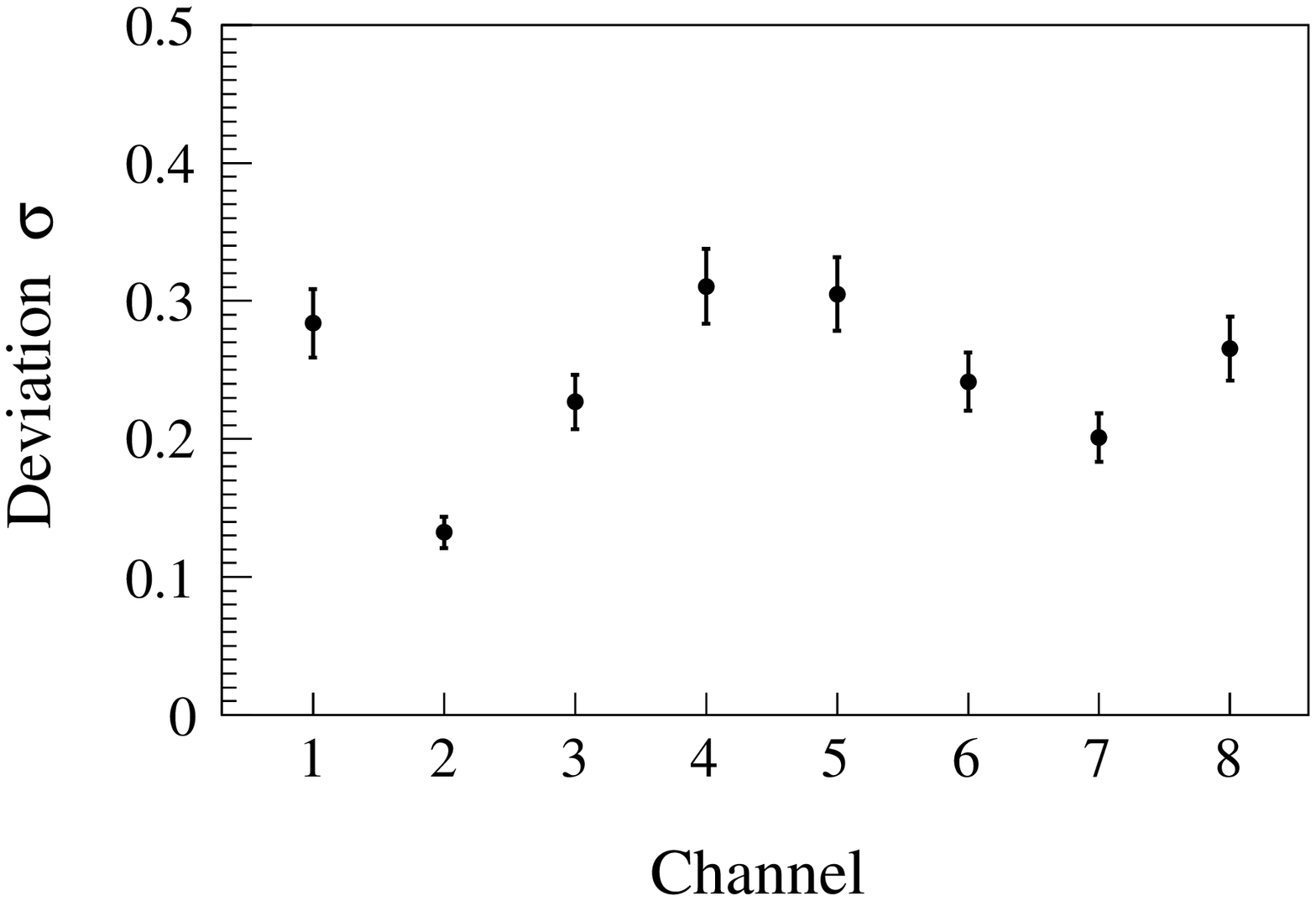}
\caption{Deviation $\sigma$ of the differential nonlinearity
obtained from Eq.~(\ref{eq:deviation}).
The values are shown for all the implemented channels.}
\label{fig:DNL_rms}
\end{figure}

As a channel-level measure of $D_i$,
the deviation $\sigma$ is defined for each channel by
\begin{equation}
\sigma^2 = \frac{1}{N}\sum_{i=0}^{N-1} D_i^2,
\label{eq:deviation}
\end{equation}
where $N$ is the total number of bins.
Figure~\ref{fig:DNL_rms}~shows the result of the measurement of $\sigma$.
The obtained value, which varies between 0.13--0.31, depends on the channel.

\section{Integral Nonlinearity}

\begin{figure*}[tbp]
  \centering
  \subfigure[Channel 1.]{
    \includegraphics[width=0.40\textwidth]{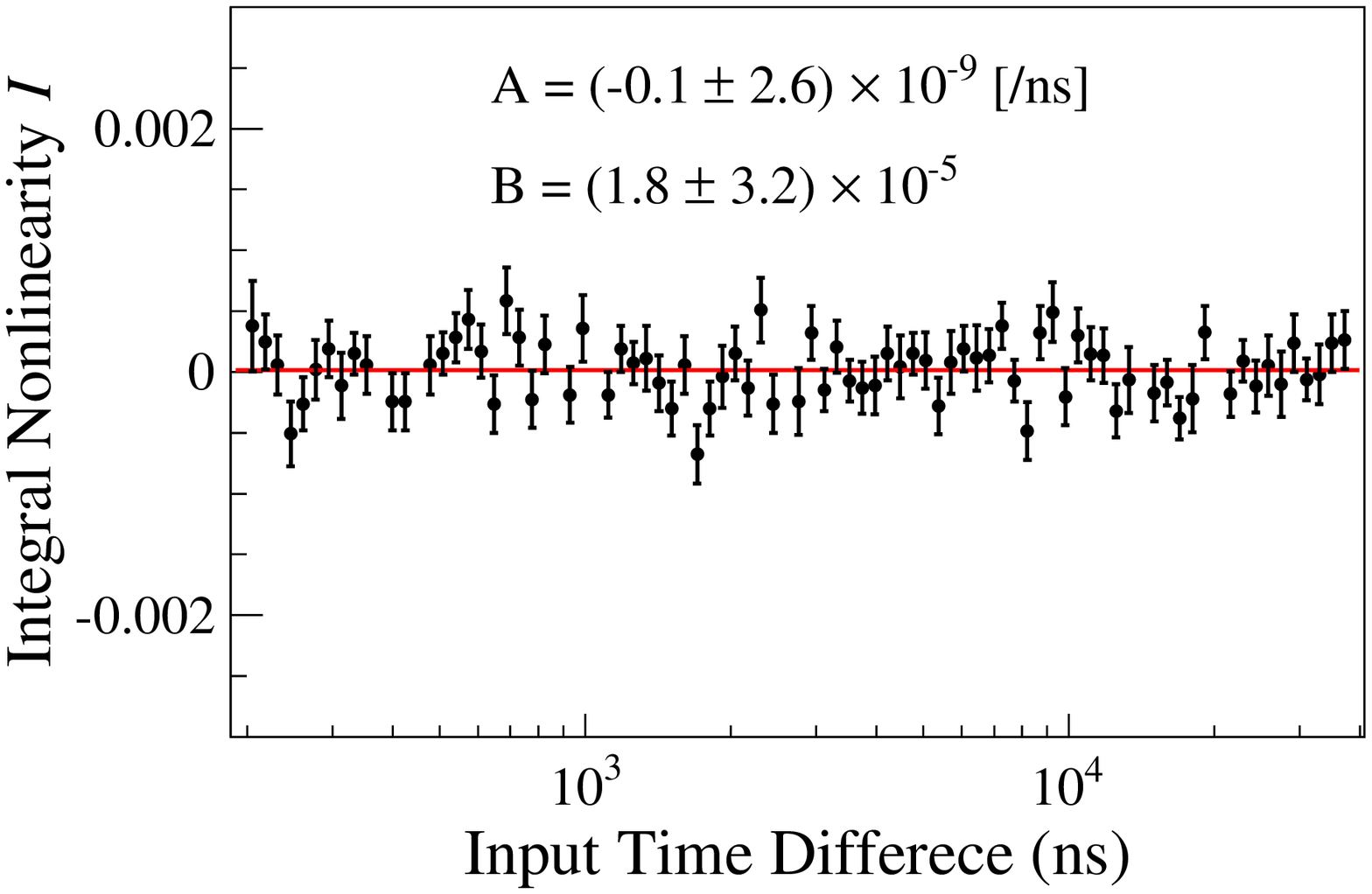}
  \label{fig:label_(a)}}
  \hspace{5mm}
  \subfigure[Channel 2.]{
    \includegraphics[width=0.40\textwidth]{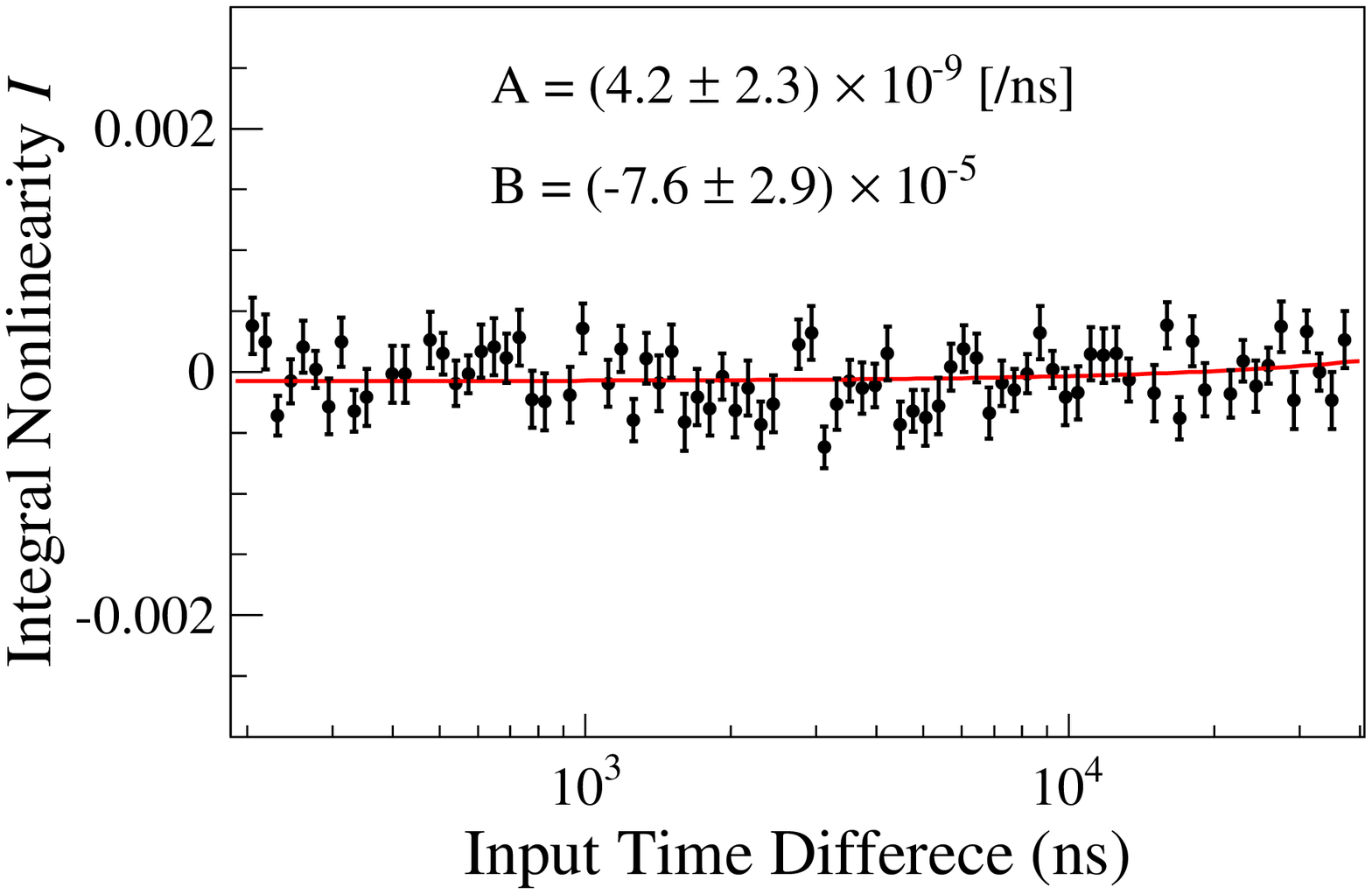}
  \label{fig:label_(b)}}\\
  \vspace{-2mm}
    \subfigure[Channel 3.]{
    \includegraphics[width=0.40\textwidth]{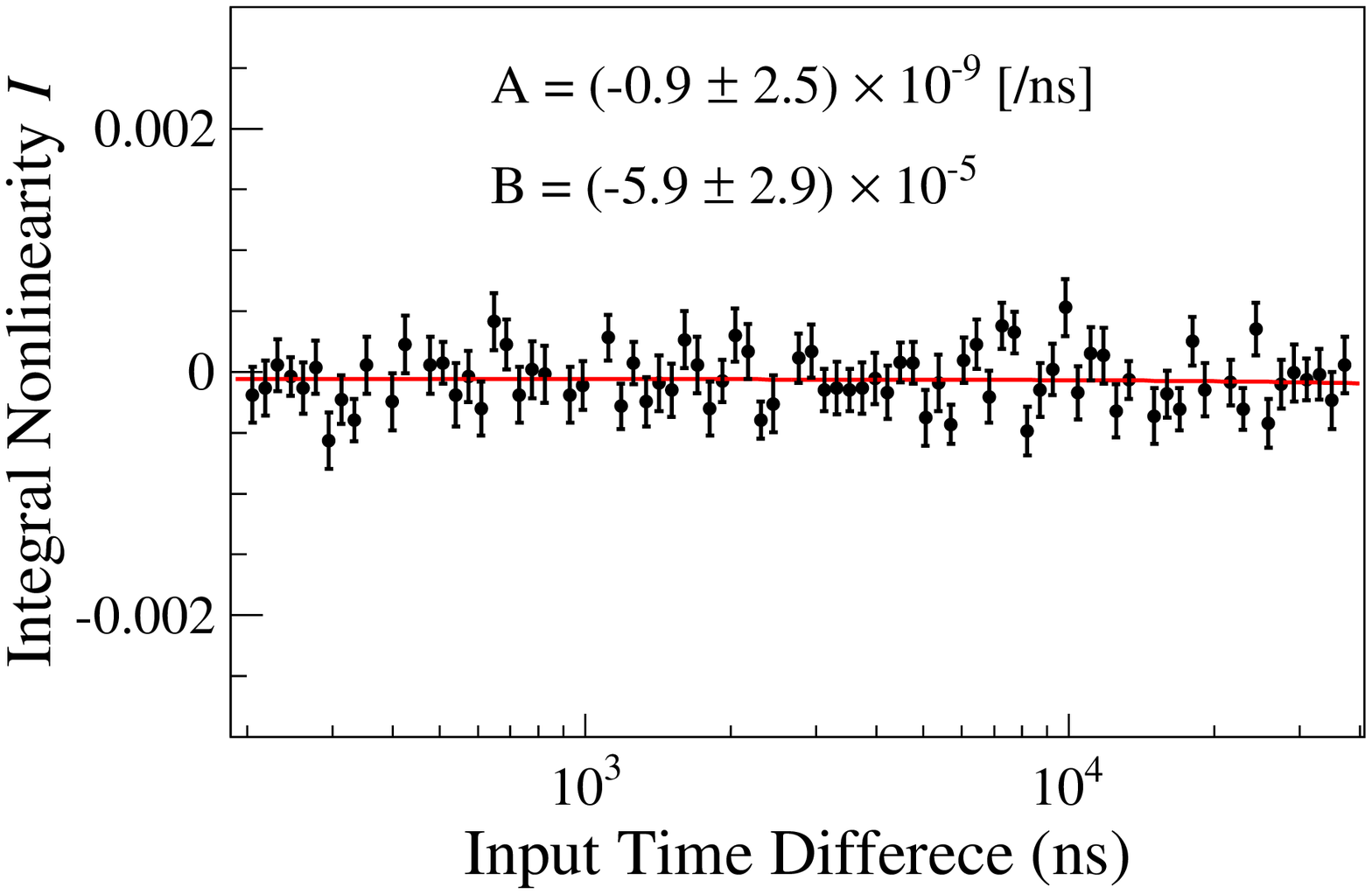}
  \label{fig:label_(c)}}
  \hspace{5mm}
    \subfigure[Channel 4.]{
    \includegraphics[width=0.40\textwidth]{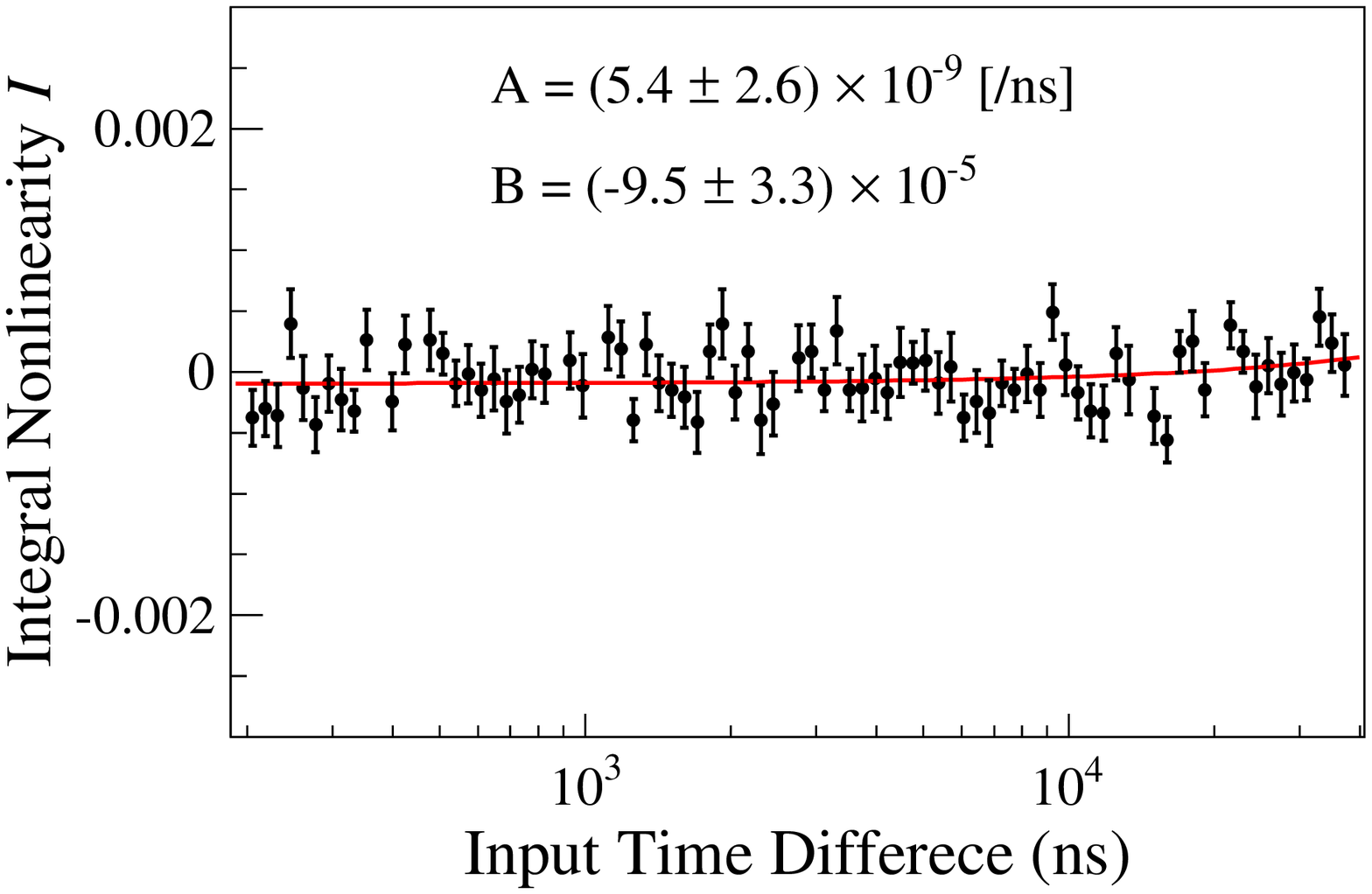}
  \label{fig:label_(d)}}\\
  \vspace{-2mm}
    \subfigure[Channel 5.]{
    \includegraphics[width=0.40\textwidth]{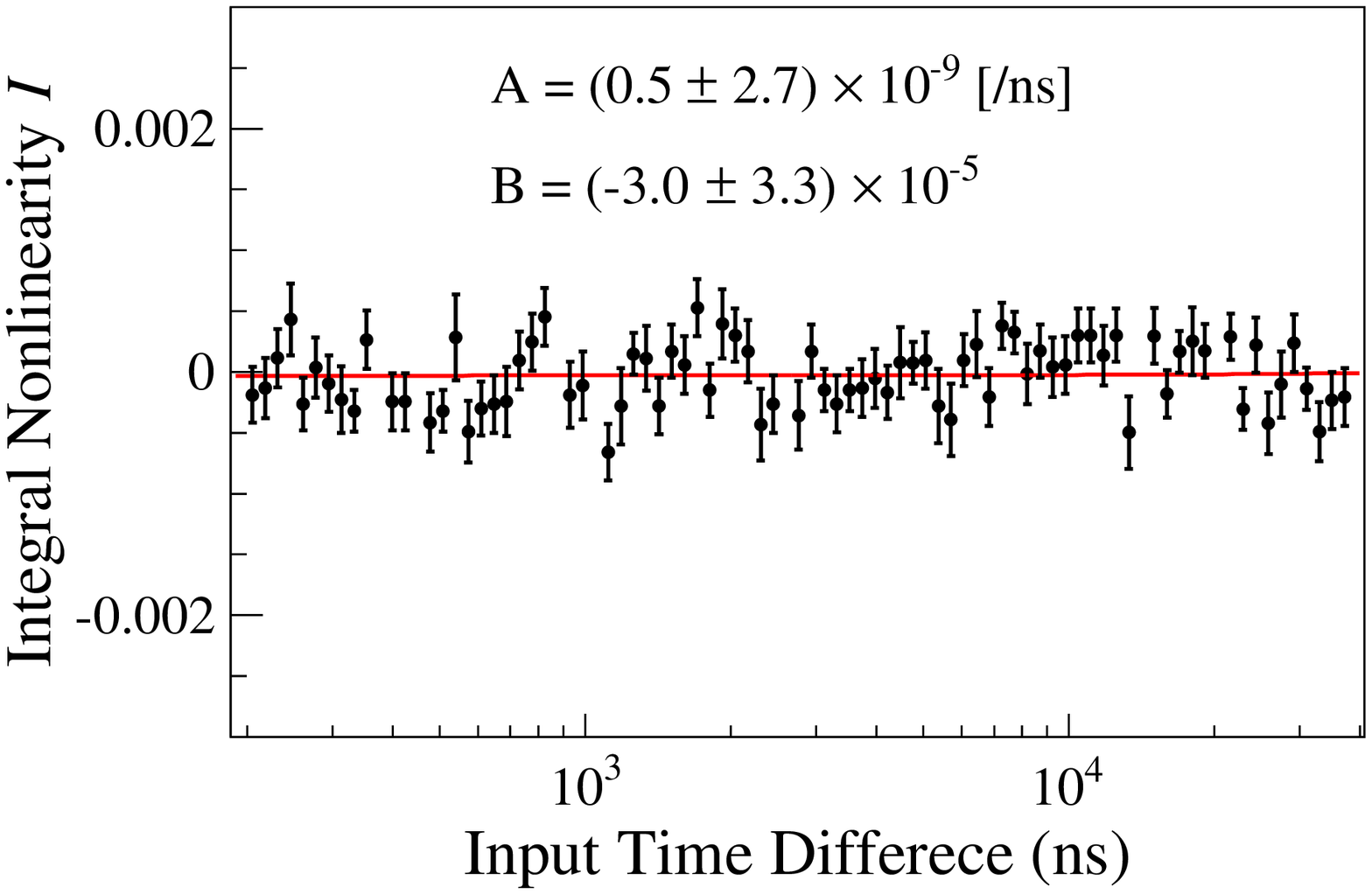}
  \label{fig:label_(e)}}
  \hspace{5mm}
    \subfigure[Channel 6.]{
    \includegraphics[width=0.40\textwidth]{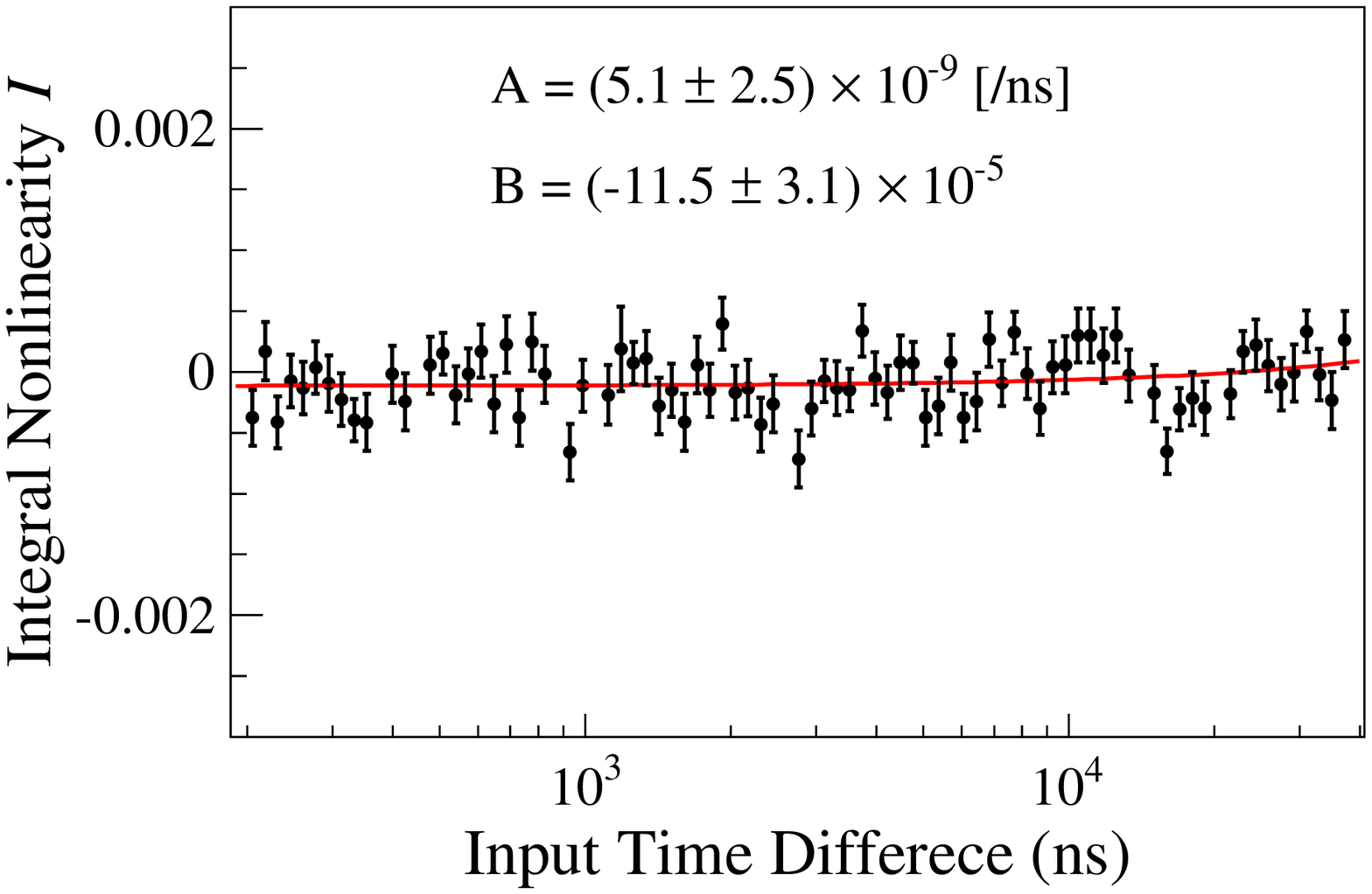}
  \label{fig:label_(f)}}\\
  \vspace{-2mm}
    \subfigure[Channel 7.]{
    \includegraphics[width=0.40\textwidth]{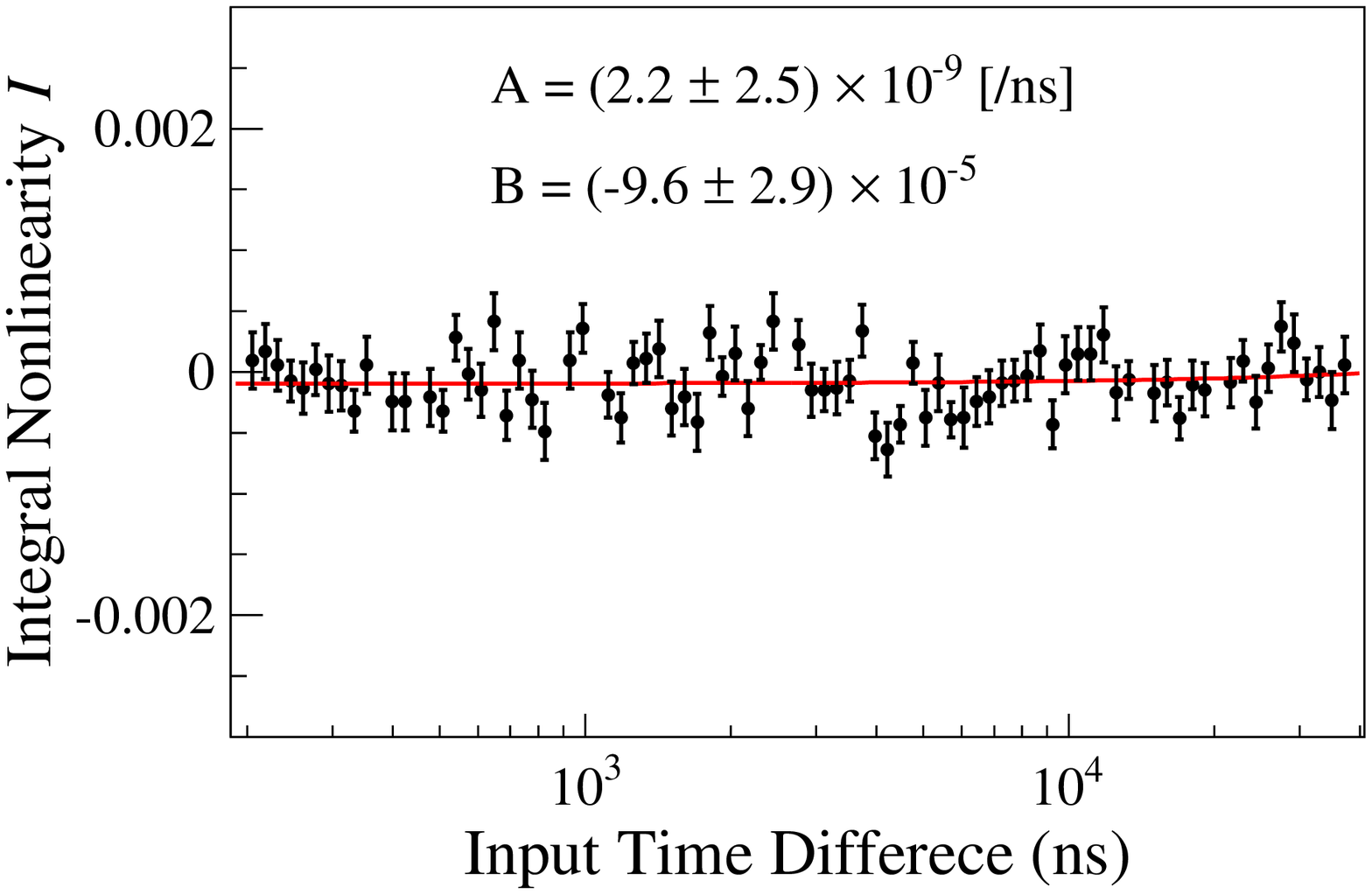}
  \label{fig:label_(g)}}
  \hspace{5mm}
    \subfigure[Channel 8.]{
    \includegraphics[width=0.40\textwidth]{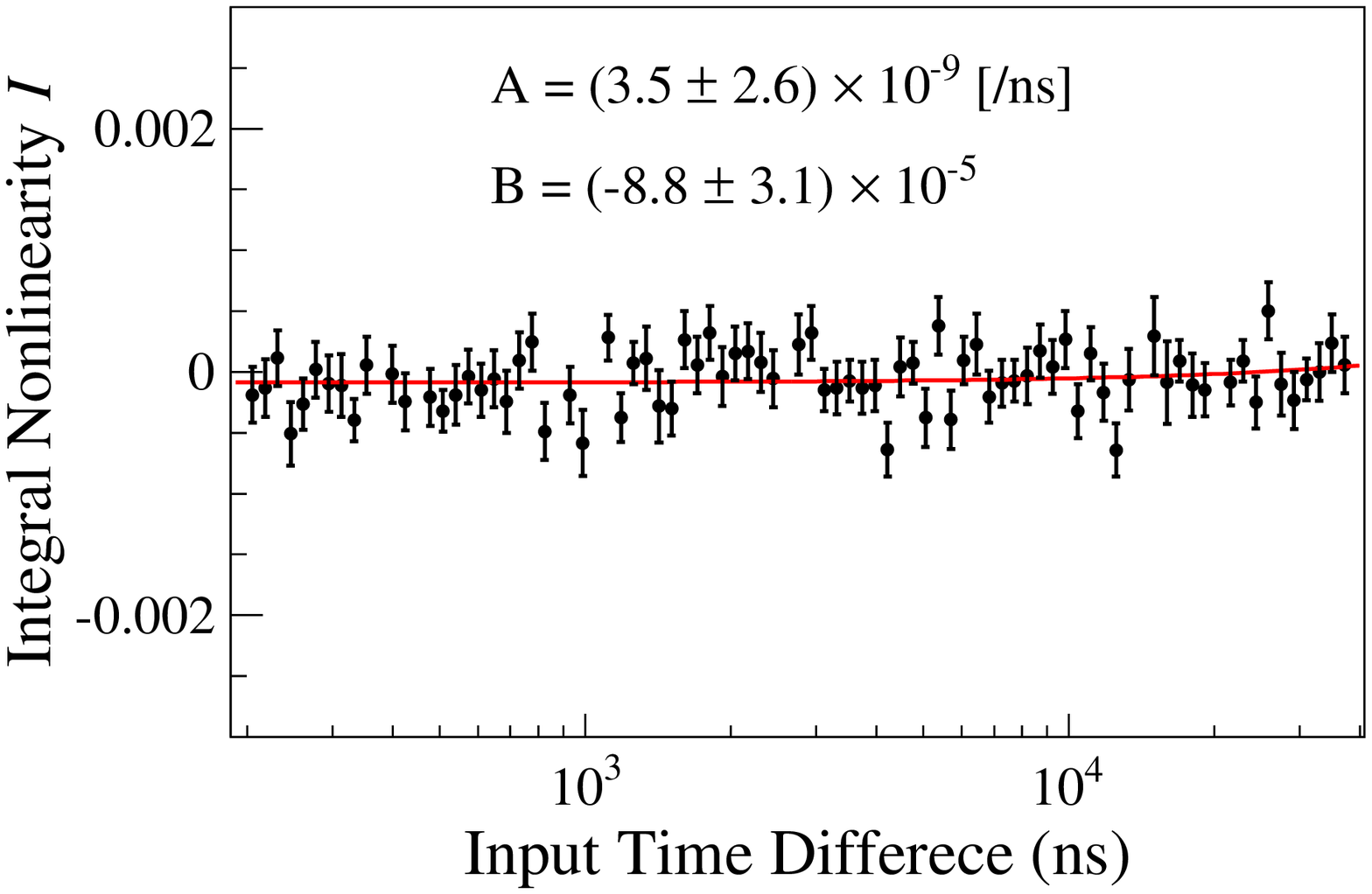}
  \label{fig:label_(h)}}
  \caption{Result of the measurement of the integral nonlinearity $I$.
  The values are shown depending on the input time difference
  for all the implemented channels 1--8.
  Curves show the results of a linear fit by $I = A \cdot T_{\rm input} + B$.
  Fitted values of parameters $A$ and $B$ are also shown for each plot.}
  \label{fig:INL}
\end{figure*}

The integral nonlinearity $I$ is given by
\begin{equation}
\label{eq:yyy}
I = \frac{\bigl< T_{\rm measured} \bigr>-T_{\rm input}}{t_{\rm bin}},
\end{equation}
where $T_{\rm input}$ is the time difference between the leading edges of the input signal clock
and $\bigl< T_{\rm measured} \bigr>$ is the mean of the measurements.
The time difference was scanned up to 37~$\mu$s,
and $\bigl< T_{\rm measured} \bigr>$ is obtained for each time difference.
The graph is fitted by a linear function $I = A \cdot T_{\rm input} + B$.
The fit result is shown in Figure~\ref{fig:INL}.
The deviation between the slope parameter $A$ and zero is at most $\sim2\sigma$,
where the uncertainty of $A$ is only 1~ps over 37~$\mu$s.
The magnitude of the mean of the offset parameter $B$ is at most $1\times10^{-4}$,
which corresponds to an offset of only 0.03~ps.

\section{Time Resolution}

\begin{figure}[tbp] 
\centering
\includegraphics[width=0.48\textwidth]{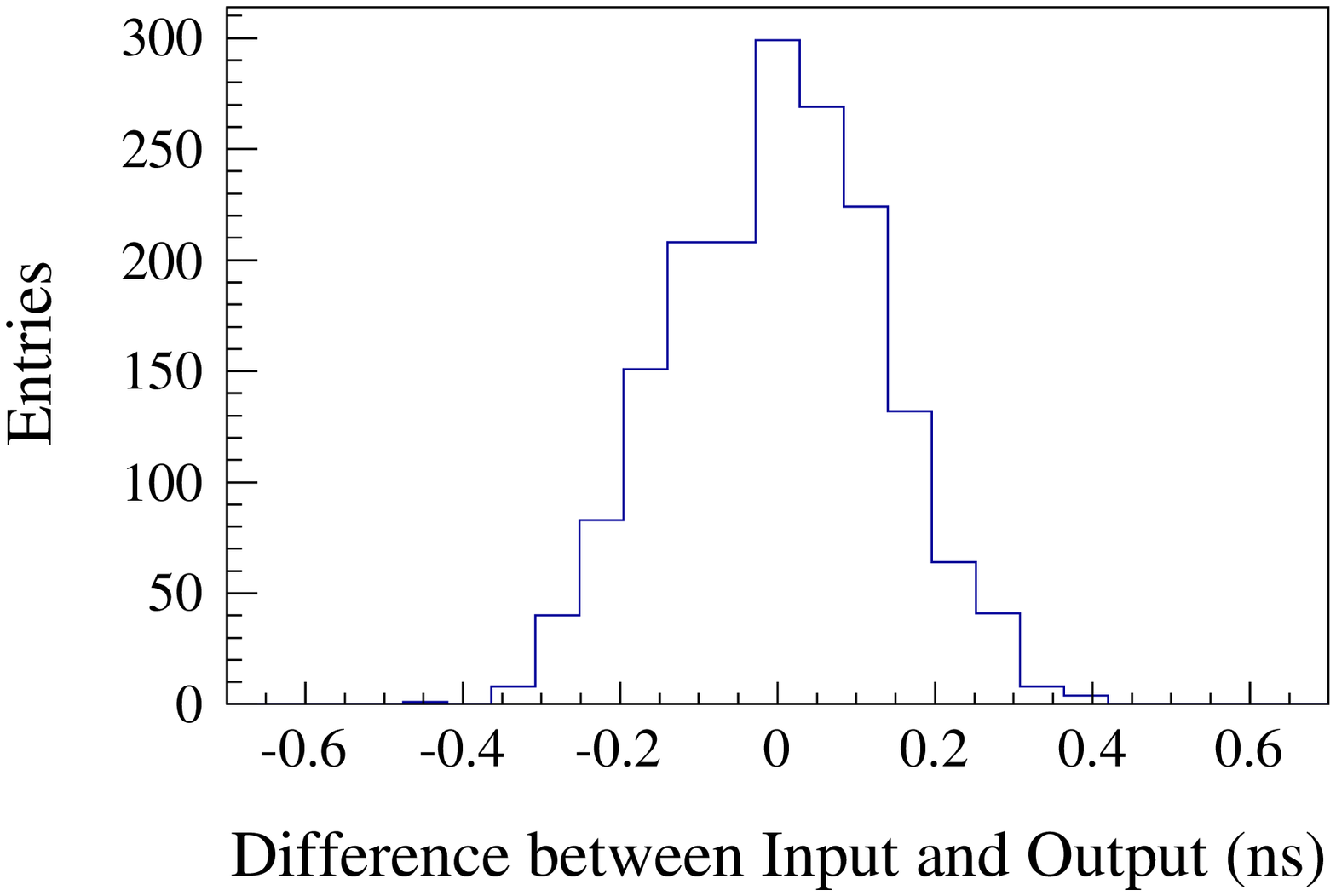}
\caption{Distribution of the difference between the measured and input time difference.
The time difference between the neighbouring leading edges of a clock is considered.}
\label{fig:resolution}
\end{figure}

\begin{figure}[tbp] 
\centering
\includegraphics[width=0.48\textwidth]{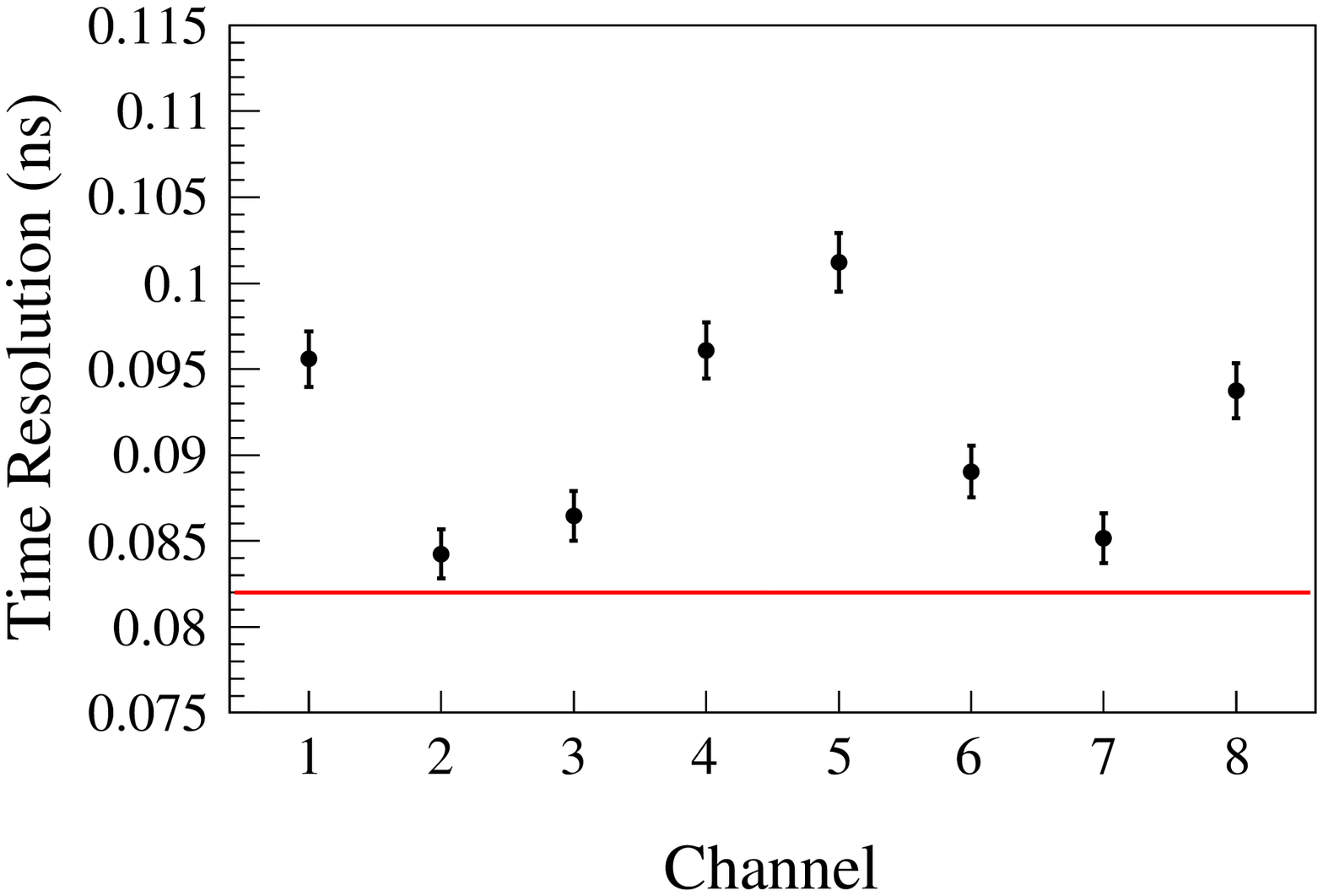}
\caption{Time resolution depending on the channels.
Line indicates the quantisation error.
The deviation from the quantisation error arises from the nonlinearity of the TDC.}
\label{fig:resolution_channel}
\end{figure}

The time resolution was evaluated 
from the measurement of the time difference between the neighbouring leading edges
of the input signal clock.
The cycle of the input signal clock was scanned from 200~ns to 37~$\mu$s.
Figure~\ref{fig:resolution} shows the difference between the measured and input time difference.
Figure~\ref{fig:resolution_channel} shows the standard deviation of the difference for each channel.
The standard deviation is considered to be the time resolution.
The obtained time resolution depends on the channel and ranges 0.08--0.10~ns.
The linear correlation factor between the time resolution
and the deviation $\sigma$ shown in Figure~\ref{fig:DNL_rms} is 0.9.
This result indicates that the time resolution depends
on the difference between the input signal paths.

For a bin size of 0.28~ns, the time resolution is larger
than the quantisation error as shown in Figure~\ref{fig:resolution_channel}.
The difference between the input signal paths has no dependence on the bin size.
On the other hand, the quantisation error has a linear relation with the bin size.
Therefore, the time resolution is closer to the quantisation error
for larger bin sizes.
As an example, for a bin size of 0.78~ns,
the measured time resolution was 0.23~ns for most channels.

\section{Temperature Dependence}

\begin{figure}[tbp] 
\centering
\includegraphics[width=0.48\textwidth]{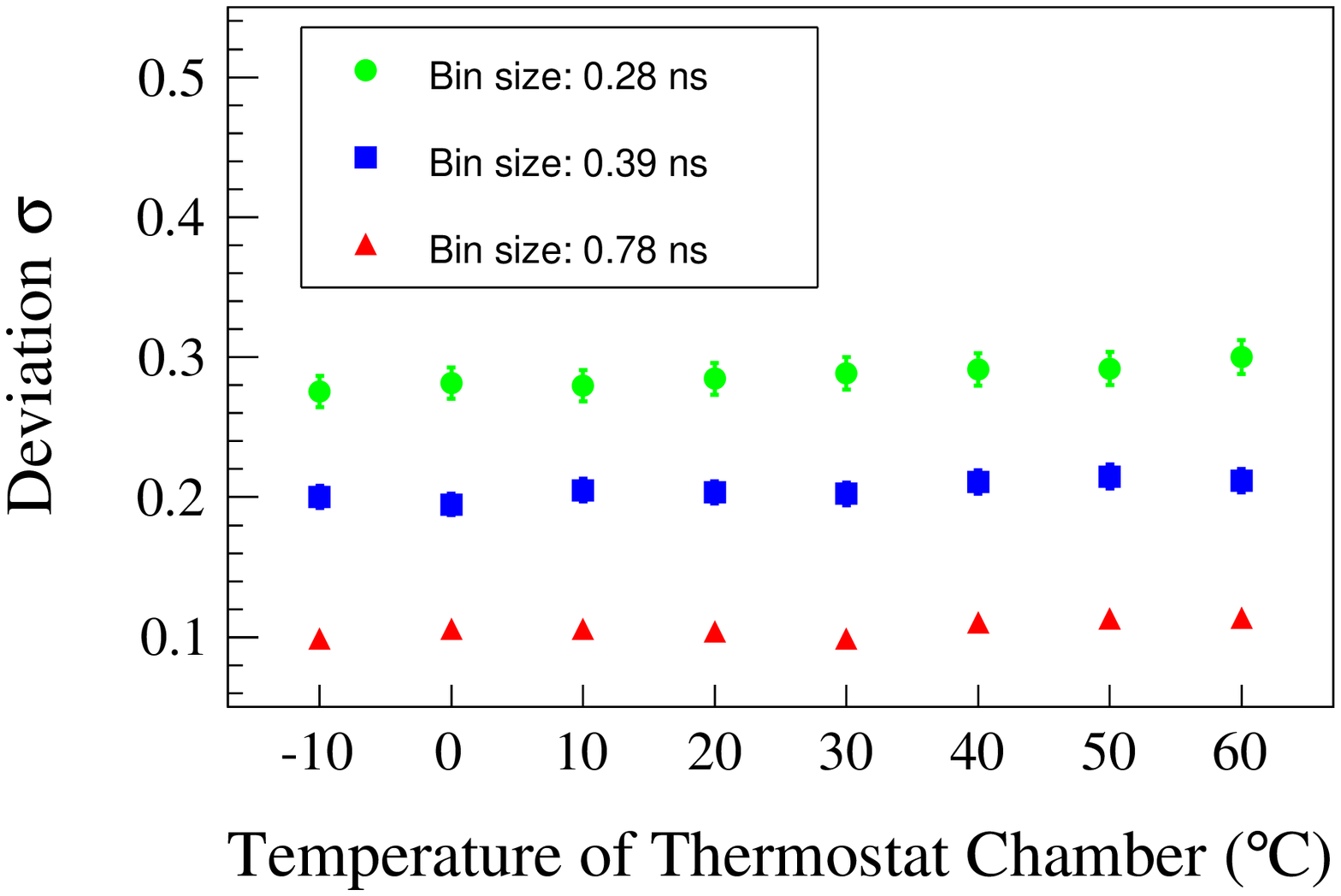}
\caption{Temperature dependence on the deviation $\sigma$
of the differential nonlinearity for the channel 8.
Circle, square, and triangle plots correspond to bin sizes of 0.28~ns, 0.39~ns, and 0.78~ns, respectively.}
\label{fig:temp_dnl}
\end{figure}

\begin{figure}[tbp] 
\centering
\includegraphics[width=0.48\textwidth]{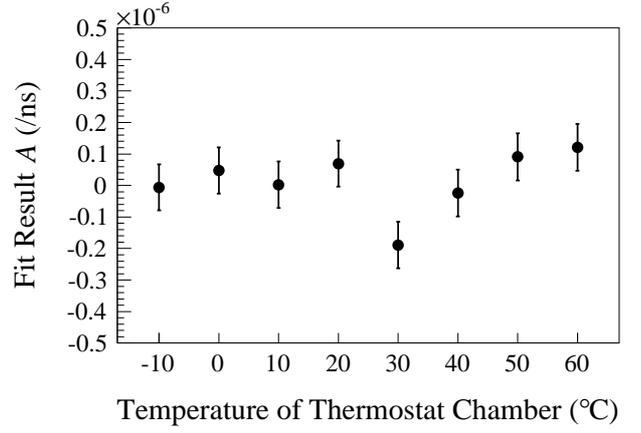}
\caption{Temperature dependence on the slope of the linear fit
for the integral nonlinearity for the channel 8.
The values are shown for a bin size of 0.28~ns.}
\label{fig:temp_inl}
\end{figure}

\begin{figure}[tbp] 
\centering
\includegraphics[width=0.48\textwidth]{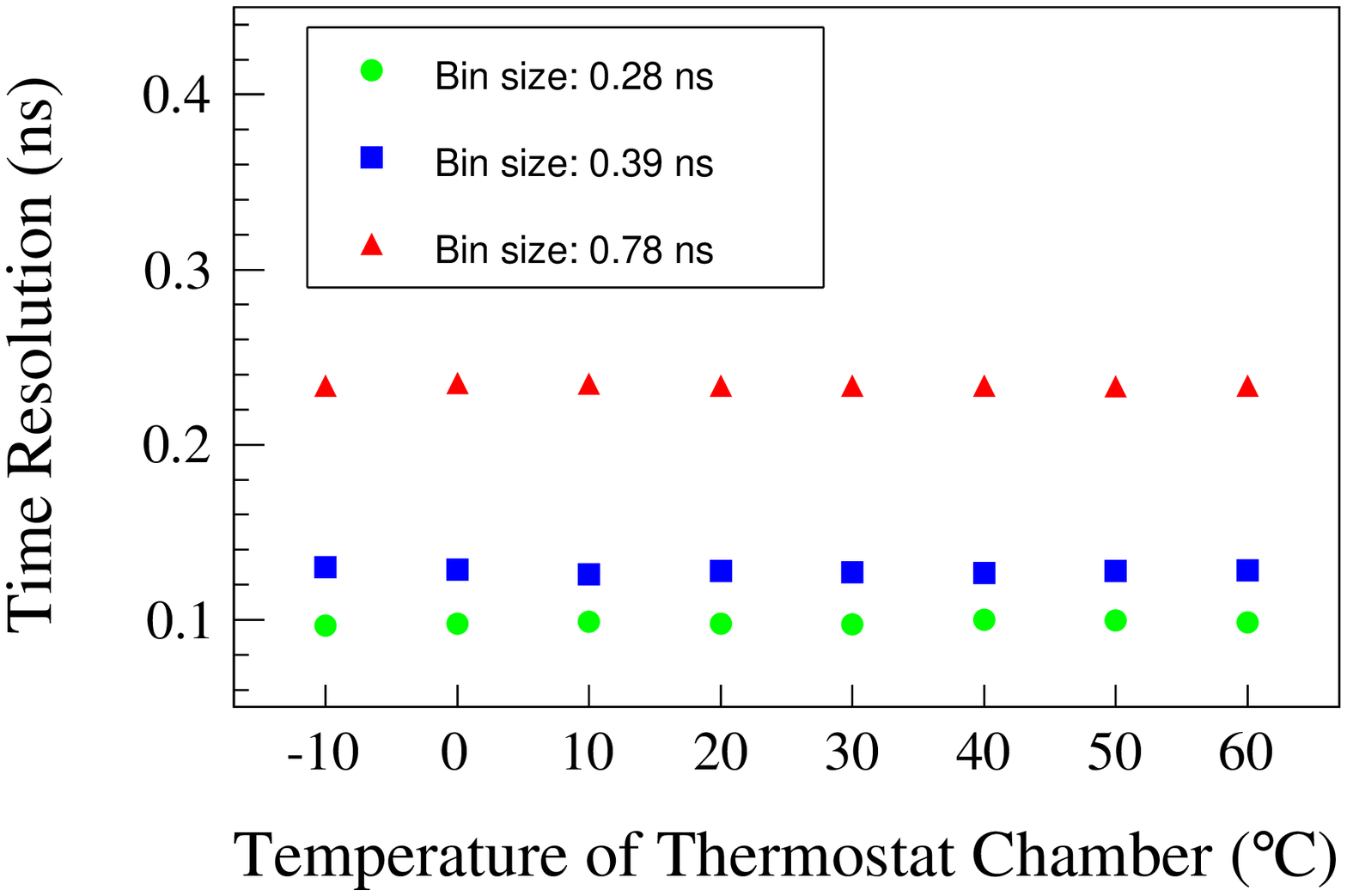}
\caption{Temperature dependence on the time resolution for the channel 8.
Circle, square, and triangle plots correspond to bin sizes of 0.28~ns, 0.39~ns, and 0.78~ns, respectively.}
\label{fig:temp_resolution}
\end{figure}

The temperature dependences of $D_i$, $I$, and the time resolution were evaluated
with a thermostat chamber ESPEC~SH-641~\cite{ESPEC}
for a range of the chamber temperature from $-10~{}^\circ$C to $60~{}^\circ$C.
The data were acquired after the temperature of FPGA became stable during the operation.
The temperature of FPGA during operation was about 8~${}^\circ$C
higher than that of the chamber.
Figure~\ref{fig:temp_dnl} shows the relation between the deviation $\sigma$ of $D_i$ and the temperature.
Figure~\ref{fig:temp_inl} shows the relation
between the slope parameter $A$ from the linear fit to $I$ and the temperature.
Figure~\ref{fig:temp_resolution} shows the relation
between the time resolution and the temperature.
The temperature dependence is small in the investigated temperature range.

\section{Power Consumption}

The power consumption of FPGA for the eight-channel TDC was estimated by the Vivado design tool.
Table~\ref{tab:TDC_power} summarises the results
on the FPGA temperature of 28~$^\circ$C and 68~$^\circ$C
for bin sizes of 0.28~ns and 0.78~ns.
Since the temperature of the FPGA during the measurement was 8~${}^\circ$C
higher than that of the thermostat chamber, we took 28~$^\circ$C and 68~$^\circ$C as the references.
The dominant contributors for the dynamic power consumption are
found to be the clock manager and SiTCP, which spend about 40\% and 30\%, respectively. 
Assuming a linear relation between the number of channels and the power consumption,
the power consumption per channel for the bin sizes of 0.28~ns and 0.78~ns
is 0.02~W and 0.01~W, respectively.

\begin{table}[tbp]
\small
\caption{Power consumption of the FPGA estimated by the Xilinx design tool.
The temperature of the FPGA during the power consumption measurement was 8~${}^\circ$C
higher than that of the thermostat chamber.
Therefore, we employ 28~$^\circ$C and 68~$^\circ$C
for a comparison with the measurement
based on the temperature of the thermostat chamber of 20~$^\circ$C and 60~$^\circ$C, respectively.}
\label{tab:TDC_power}
\smallskip
\centering
\begin{tabular}{|c|c|c|c|c|}
\hline
Bin size & FPGA temperature & Dynamic & Static & Total\\
$[\rm ns]$ & [$^\circ$C] & [W] & [W] & [W]\\ \hline \hline
0.28 & 28 & 0.56 & 0.18 & 0.73\\  \hline
0.28 & 68 & 0.56 & 0.59 & 1.15\\  \hline
0.78 & 28 & 0.43 & 0.18 & 0.61\\   \hline
0.78 & 68 & 0.43 & 0.59 & 1.02\\  \hline
\end{tabular}
\end{table}

\begin{figure}[tbp] 
\centering
\includegraphics[width=0.48\textwidth]{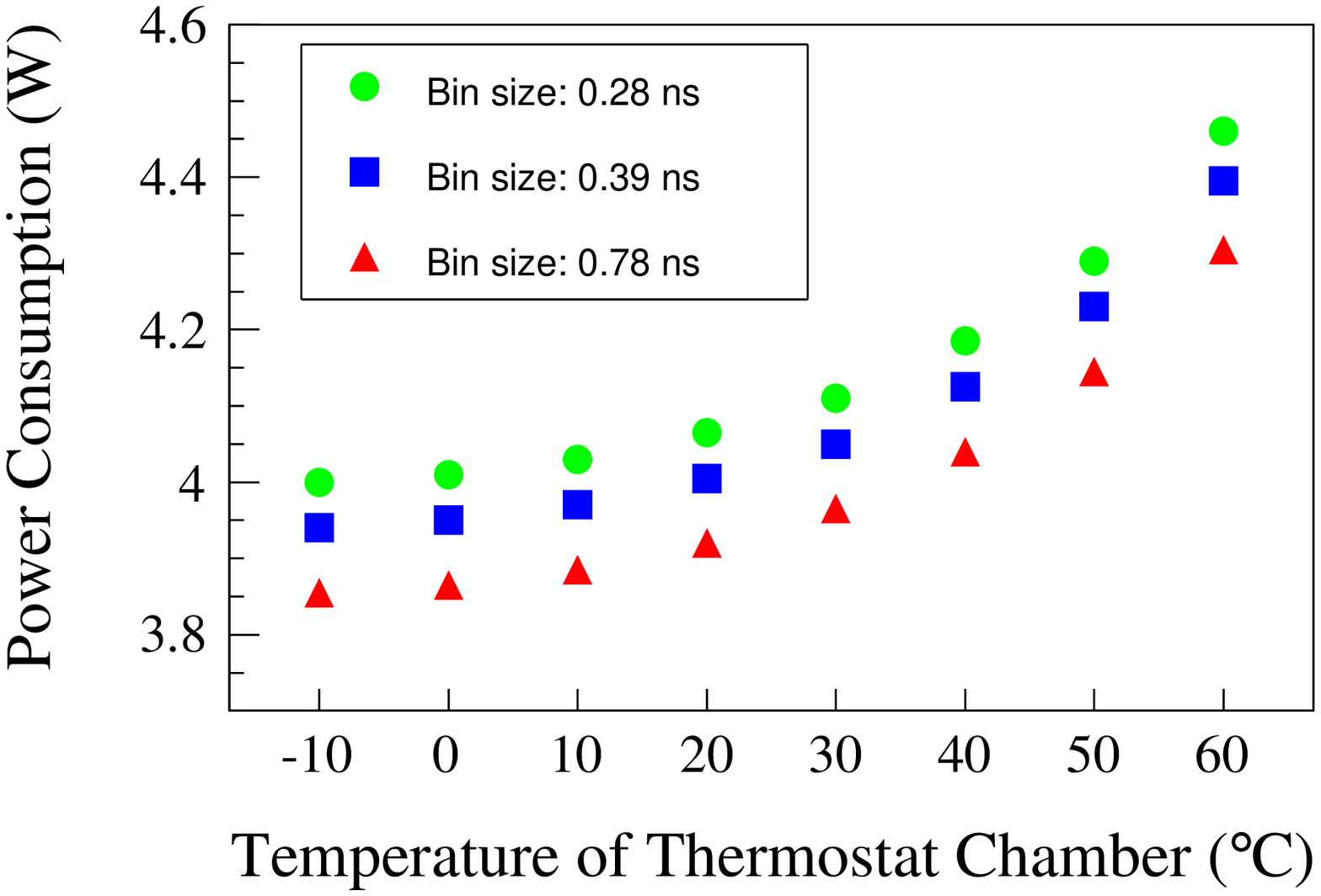}
\caption{Measured power consumption of the demonstrator during the operation.
Circle, square, and triangle plots correspond to bin sizes of 0.28~ns, 0.39~ns, and 0.78~ns, respectively.}
\label{fig:PT7_power}
\end{figure}

Figure~\ref{fig:PT7_power} shows the measured power consumption of the demonstrator.
The measured power consumption for the 0.28-ns bin size
at a chamber temperature 20~$^\circ$C is 4.1~W.
The static power consumption of the demonstrator for the same bin size
and the same temperature measured
without firmware implemented in FPGA is 2.0~W.
The dynamic power consumption of Texas Instruments DP83865DVH~\cite{Texas} for Ethernet data transfer,
which is considered the main consumer on the demonstrator other than FPGA,
is estimated to be~1.3 W.
The value of $4.1~{\rm W} - (2.0~{\rm W} + 1.3~{\rm W}) = 0.8~{\rm W}$ is
close to the value estimated from the Vivado design tool of 0.73~W.

\section{Resources and Number of Channels}

Table~\ref{tab:resources} summarises
the utilisation of the FPGA resources for the tested eight-channel TDC.
Because multiple clocks with different frequencies are generated,
the ratio of the utilised clocking resources to the total clocking resources is relatively large.
Most of the resources of the look-up tables, the registers, and the random-access memory are spent by SiTCP.

\begin{table}[tbp]
\small
\caption{Utilisation of the FPGA resources for eight-channel TDC on the Kintex-7 FPGA.
The available resources and the ratios of the utilised and available resources are also shown.
The ratio is the highest for the clocking resource.}
\smallskip
\centering
\begin{tabular}{|c|c|c|c|}
\hline
 Resource & Utilised & Available & Ratio [\%]\\ \hline\hline
 Look-up tables & 4361 & 203800 & 2.1 \\ \hline
 Registers & 5939 & 407600 & 1.5 \\ \hline
 Memory & 48 & 445 & 10.8 \\ \hline
 Input/output ports & 60 & 582 & 10.3 \\ \hline
 Clocking & 14 & 32 & 43.8 \\ \hline
\end{tabular}
\label{tab:resources}
\end{table}

We developed firmware with 256 channels
to discuss the potential extension of the number of channels.
Table~\ref{tab:resources_256} shows the utilised resources.
The ratio of the utilised input/output ports to the total input/output ports is 58\%,
while the ratios for the other elements are smaller.
The difference in the divided input signal delay for each channel
was evaluated by the Vivado design tool.
The standard deviation of the input signal delay is 0.14~ns.
The clock skew may affect the offset between the channels,
but it is within a range supported for the employed Kintex-7 FPGA.

\begin{table}[tbp]
\small
\caption{Utilisation of the FPGA resources for 256-channel TDC on the Kintex-7 FPGA.
The available resources and the ratios of the utilised and available resources are also shown.
The ratio is the highest for the input/output ports.}
\smallskip
\centering
\begin{tabular}{|c|c|c|c|}
\hline
 Resource & Utilised & Available & Ratio [\%]\\ \hline\hline
 Look-up tables & 27682 & 203800 & 13.6 \\ \hline
 Registers & 48826 & 407600 & 12.0 \\ \hline
 Memory & 180 & 445 & 40.4 \\ \hline
 Input/output ports & 292 & 582 & 58.4 \\ \hline
 Clocking & 14 & 32 & 43.8 \\ \hline
\end{tabular}
\label{tab:resources_256}
\end{table}

\section{Conclusion}

An eight-channel TDC with a variable bin size down to 0.28~ns
was implemented in a Xilinx Kintex-7 FPGA (XC7K325T-2FFG900)
and tested.
The TDC is based on a multisampling scheme
with quad phase clocks
synchronised with an external reference clock.
The measured time resolution for a 0.28-ns bin size is 0.08--0.10~ns,
depending on the channel.
The performance has a negligible dependence on the temperature.
Although the actual performance was not demonstrated,
it is possible to provide the firmware of a 256-channel TDC
in a Xilinx design tool
with a 0.14~ns standard deviation of the input signal delay.

Implementation of TDCs in FPGAs increases the flexibility to modify logics.
This feature is advantageous
when coping with the changes in the experimental conditions.
If a stable reference clock is available such as in high-energy physics experiments,
the TDC in this study does not need to be calibrated.
In addition to the flexibility, FPGAs have advantages
on the availability of various highly reliable circuits,
e.g. clock managers and data transceivers.
Consequently, the FPGA-based TDCs are useful
in fields requiring subnanosecond resolution and simple and robust operation.

\section*{Acknowledgements}

We acknowledge the support of the demonstrator developers
Hiroshi Sakamoto, Chikuma Kato, Takayuki Tokunaga,
and Yusaku Urano, at The University of Tokyo, Tokyo, Japan.
We are also grateful for the support from the Open-It Consortium.


\end{document}